\PassOptionsToPackage{math-style=ISO}{unicode-math}
\documentclass[11pt,british]{article}

\usepackage[T1]{fontenc}

\usepackage[french,italian,main=british]{babel}
\frenchsetup{ThinColonSpace=true}
\usepackage[autostyle,english=british]{csquotes}

\usepackage[a4paper]{geometry}

\usepackage[dvipsnames]{xcolor} %
\usepackage{graphicx} %
\usepackage{titling} %
\usepackage[mode=buildmissing]{standalone} %
\usepackage[inline,shortlabels]{enumitem} %
\usepackage{subcaption} %
\usepackage{pdfpages} %

\usepackage{amsmath}
\usepackage{amssymb} %
\usepackage{mathtools}

\usepackage[hidelinks]{hyperref} %
\hypersetup{%
    pdfauthor={A. A. Rispo Constantinou, B. Magyari, G. Ianniruberto, E. van Ruymbeke, D. J. Read},
    pdftitle={Entropic phase separation in vitrimer–polymer melts},
    pdfsubject={Vitrimer physics preprint},
    pdfkeywords={vitrimers, vitrimer blends, entropy-driven, phase separation, associative covalent adaptive networks (asso-CANs), dynamic polymer networks (DPNs), material design},
    pdfproducer={LaTeX},
    pdfcreator={pdflatex},
    }
\usepackage{cleveref}

\usepackage[%
    autocite=inline,%
    backend=biber,%
    style=ext-authoryear-ecomp,
    doi=true,%
    articlein=false,
    maxcitenames=2,
    minbibnames=3,
    maxbibnames=4,
    ]{biblatex}
\renewbibmacro*{doi+eprint+url}{%
    \printfield{doi}%
    \newunit\newblock%
    \iftoggle{bbx:eprint}{%
        \usebibmacro{eprint}%
    }{}%
    \newunit\newblock%
    \iffieldundef{doi}{%
        \usebibmacro{url+urldate}}%
        {}%
    }

\usepackage[raise]{fmtcount} %
\DefineBibliographyExtras{british}{%
	
}
\usepackage[useregional,en-GB]{datetime2} \DTMlangsetup[en-GB]{ord=raise}

\usepackage{booktabs}
\usepackage{microtype} %
\usepackage[avoid-all]{widows-and-orphans}

\usepackage[p]{libertinus-type1}
\usepackage[lcgreekalpha,ISO]{libertinust1math}

\usepackage{tikz}
\usetikzlibrary{shapes}
\usetikzlibrary{calc} %
\usetikzlibrary{through} %
\usetikzlibrary{decorations.pathmorphing}
\usetikzlibrary{hobby}
\usepackage{pgfplots}
\pgfplotsset{compat=1.18}
\usepgfplotslibrary{ternary,colorbrewer,colormaps,fillbetween}

\definecolor{fd}{named}{lightgray}

\definecolor{A}{named}{CornflowerBlue}
\definecolor{B}{named}{blue}
\definecolor{C}{named}{green} %
\definecolor{D}{named}{magenta}		%
\tikzstyle{as A}=[A, smooth, hobby, opacity=0.5, line width=3, line 
cap=round, line join=round]
\tikzstyle{as B}=[B, smooth, hobby, line width=3, line cap=round, line 
join=round]
\tikzstyle{as C}=[regular polygon, regular polygon sides=6, fill=C, scale=0.6, draw, thick]
\tikzstyle{as D}=[regular polygon, regular polygon sides=6, fill=D, scale=0.6, draw, thick]
\tikzstyle{line CD}=[very thick, line cap=rect]

\definecolor{stablesim}{named}{ForestGreen}
\definecolor{dubiousim}{named}{Bittersweet}
\definecolor{separasim}{named}{RubineRed}

\usepackage{chemfig}
\setchemfig{debug=false, bond join=true, atom sep=1.5em}

\pgfplotsset{grid style={ultra thin}}

\newcommand{\drawsims}[4]{%
	\ifx?#2? \else%
	\foreach \p in {#2} {\addplot[only marks, stablesim, mark=Mercedes star flipped] coordinates {(1-#1,\p/100)};}%
	\fi%
	\ifx?#3? \else%
	\foreach \p in {#3} {\addplot[only marks, dubiousim, mark=star] coordinates {(1-#1,\p/100)};}%
	\fi%
	\ifx?#4? \else%
	\foreach \p in {#4} {\addplot[only marks, separasim, mark=10-pointed star] coordinates {(1-#1,\p/100)};}%
	\fi%
}%

\newcommand{\beaddiameter}{1.5em}

\tikzset{
    regular/.style={circle, draw, fill=B!60!black!50!white, 
    	minimum size=\beaddiameter},
    stickers/.style={circle, draw, fill=B!70!white, 
    minimum size=\beaddiameter},
    radiusline/.style={->, black, dash dot, thick,
    	},
    solidbond/.style={
    	B, very thick, %
    	},
    dashedbond/.style={
    	B, densely dashed, very thick,
    },
    oldcrosslink/.style={
    	D, very thick %
    	},
    newcrosslink/.style={
    	C, densely dashed, very thick
    	},
    arrow/.style={
    	->, black, thick, %
    	},
    largecircle/.style={draw, dotted, thick},
}

\pgfplotsset{
    cycle list/.define={graph_col}{
        [of colormap={viridis}, samples=19]
    },
}
\definecolor{customBlue}{HTML}{58e6df}
\definecolor{customRed}{HTML}{bd170b}
\definecolor{lightBlue}{RGB}{189,225,241}

\usepackage{soul}
\newcommand{\DIFadd}[1]{#1}
\newcommand{\DIFdel}[1]{}

\newcommand*{\A}{\textnormal{\kern-0.1em A}}
\newcommand*{\B}{\textnormal{B}}
\newcommand*{\C}{\textnormal{C}}
\newcommand*{\D}{\textnormal{D}}
\newcommand*{\degreepolym}[1]{\textnormal{dp}_{#1}}

\crefname{equation}{eq.\kern-0.1em}{eqs.\@}
\Crefname{equation}{Equation}{Equation}
\crefname{figure}{fig.\kern-0.1em}{figs.\@}
\Crefname{figure}{Figure}{Figures}

\makeatletter
\newcommand{\nb}{\em n.\,b\@ifnextchar.{\em}{.\@\em}}
\newcommand{\eg}{\em e.\,g\@ifnextchar.{\em}{.\@\em}}
\newcommand{\ie}{\em i.\,e\@ifnextchar.{\em}{.\@\em}}
\newcommand{\etc}{\em etc\@ifnextchar.{\em}{.\@\em}}
\newcommand{\cf}{\em c.\,f\@ifnextchar.{\em}{.\@\em}}
\newcommand{\wlofg}{w.\,l.\,o.\,g\@ifnextchar.{}{.\@}}
\newcommand{\qv}{\em q.\,v\@ifnextchar.{\em}{.\@\em}}
\newcommand{\qqv}{\emph{qq.\,v.}}
\newcommand{\vs}{\emph{vs}}
\newcommand{\stat}{s.\,t\@ifnextchar.{}{.\@}}
\newcommand{\lhs}{l.-h.\,s\@ifnextchar.{}{.\@}}
\newcommand{\rhs}{r.-h.\,s\@ifnextchar.{}{.\@}}
\newcommand{\Nb}{\em N.\,b\@ifnextchar.{\em}{.\@\em}}
\newcommand{\Eg}{\em E.\,g \@ifnextchar.{\em}{.\@\em}}
\newcommand{\Ie}{\em I.\,e\@ifnextchar.{\em}{.\@\em}}
\newcommand{\Cf}{\em C.\,f\@ifnextchar.{\em}{.\@\em}}
\newcommand{\Wlofg}{W.\,l.\,o.\,g\@ifnextchar.{}{.\@}}
\newcommand{\Stat}{S.\,t\@ifnextchar.{}{.\@}}
\makeatother

\newcommand*{\ds}{\ \ } %
\ifPDFTeX \let\symup\mathrm \fi
\newcommand*{\ec}{\ensuremath{\symup{e}}} %
\newcommand*{\pc}{\ensuremath{\symup{\pi}}} %
\newcommand*{\chg}{\ensuremath{\symup{\delta}}} %
\newcommand*{\Chg}{\ensuremath{\symup{\Delta}}} %
\newcommand*{\bigO}{\ensuremath{\mathcal{O}}} %
\let\oldvec\vec
\let\vec\relax
\newcommand*{\vec}[1]{\oldvec{#1}\kern0.1em}

\usepackage{authblk}
\usepackage{orcidlink}
\providecommand{\orcidlinki}[2]{#1\,\orcidlink{#2}}

\author[1,2]{\orcidlinki{Alexandros~A.~Rispo~Constantinou}{0009-0007-2960-1589}*}
\author[1,3]{\orcidlinki{Bálint~Magyari}{0009-0007-2282-8043}}
\author[1]{\orcidlinki{Giovanni~Ianniruberto}{0000-0002-6963-8753}}
\author[2]{\orcidlinki{Evelyne~van~Ruymbeke}{0000-0001-7633-0194}}
\author[4]{\orcidlinki{Daniel~J.~Read}{0000-0003-1194-9273}}

\affil[1]{DICMaPI, Università degli Studî di Napoli Federico~II, 80125~Naples, Italy}
\affil[2]{BSMA, IMCN, Université catholique de Louvain, 1348~Louvain-la-Neuve,
Belgium}
\affil[3]{MSE, University of Crete, 70013~Heraklion, Greece}
\affil[4]{School of Mathematics, University of Leeds, Leeds~LS2\,9JT, United Kingdom}

\title{Entropic Phase Separation in Polymer--Vitrimer Melts}
\date{%
	*E-mail:
	\href{mailto:alexandrosandreas.constantinou@unina.it}
	            {alexandrosandreas.constantinou@unina.it}\\[1em]
	\DTMdate{2026-07-21}
}

\begin{document}
\maketitle

\IfFileExists{./toc-graphic.pdf}{}{%
	\immediate\write18{%
		pdflatex -shell-escape -interaction=nonstopmode toc-graphic.tex
	}
}
\IfFileExists{./fig-anatomy.pdf}{}{%
	\immediate\write18{%
		pdflatex -shell-escape -interaction=nonstopmode fig-anatomy.tex
	}
}
\IfFileExists{./fig-BERs.pdf}{}{%
	\immediate\write18{%
		pdflatex -shell-escape -interaction=nonstopmode fig-BERs.tex
	}
}

\begin{center}
	\includegraphics{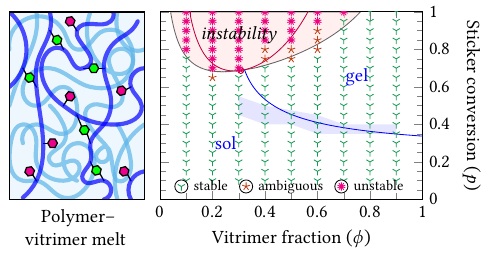}
\end{center}

\begin{abstract}
Traditional plastics demand a choice between durability (thermosets) and
reprocessability (thermoplastics). Vitrimers are a recent class of polymer
network combining both these qualities. Their increased cost of production
can be offset by mixing them with a traditional thermoplastic; however,
phase separation in such blends can lead to inhomogenous materials.
In this paper, we adapt an existing model for the free energy of dissociative
polymer networks to their associative, vitrimeric counterpart. We test the
accuracy of the model's predictions by comparing them with the results of
novel molecular-dynamics simulations.
We demonstrate that such melts can undergo phase separation even in the absence
of energetic interactions between the components, \DIFdel{and that
that the}\DIFadd{with a} phase diagram \DIFdel{of the melts is}qualitatively similar to that of
dissociative systems. \DIFadd{We furthermore find that --- for equal-length
chains in the limit of highly functionalised vitrimers --- entropic phase
separation sets in as soon as two function sites per chain are cross-linked to
other chains, such that only loosely cross-linked networks are stable}.
\end{abstract}

\begin{description}\small
    \item[\textbf{Keywords}] vitrimers, polymer melts, entropy-driven
    phase separation, associative covalent adaptive networks (asso-CANs),
    dynamic polymer networks (DPNs), \DIFadd{bifunctional cross-links}
\end{description}

\clearpage~\vfill
\begin{otherlanguage}{french}
	\begin{abstract}
        Traditionnellement, le plastique était soit durable, soit recyclable:
        soit thermodurci, soit thermoplastique. Les vitrimères sont
        une récente catégorie de matériau unissant ces deux qualités. Leur coût de
        production plus élévé peut être compensé en les mêlant à un polymère
        thermoplastique; cependant, les mélanges vitrimère/polymère sont susceptibles
        à la démixtion, ainsi présentant un obstacle à l'obtention de mélanges homogènes.
		Cet article prend un modèle pré-existant pour l'energie libre de réseaux
        polymériques dissociatifs et l'adapte au cas vitrimérique, à réticulation
        associative. Afin d'en évaluer la fiabilité, les prédictions du modèle sont
        ensuite comparées aux résultats de nouvelles simulations de dynamique
        moléculaire. Nous constatons que les mélanges vitrimère/polymère peuvent
        subir une séparation de phase même sans interactions énergétiques entre
        leurs composantes.  De plus, le diagramme de phases obtenu théoriquement
        est semblable à celui de systèmes dissociatifs, avec la particularité que
        \DIFadd{--- pour des chaînes de longueurs égales et quand le nombre de
        fonctions par chaîne est grand --- la démixtion a lieu dès que deux sites
        par chaîne vitrimérique sont crochés à d'autres chaînes.}
	\end{abstract}
    \begin{description}\small
    \item[\textbf{Mots-clef}] vitrimères, mélanges de polymères, séparation de phase,
        démixtion entropique, réseaux covalents adaptables associatifs,
        réticulation polymérique dynamique, \DIFadd{liens bifaces}
    \end{description}
\end{otherlanguage}
\vfill
\begin{otherlanguage}{italian}
	\begin{abstract}
		Le plastiche tradizionali esigono una scelta tra durabilità (polimeri
        termoindurenti) e riciclabilità (termoplastici). I vitrimeri, invece, sono
        una recente classe di materiali che conciliano entrambe le qualità. Il loro
        maggiore costo di produzione puo' essere compensato immettendoli in una
        matrice termoplastica tradizionale; tuttavia, lo smiscelamento spontaneo
        per via di separazione di fase è un ostacolo importante alla preparazione
        di miscele omogenee vitrimero-polimeriche. %
        In questo articolo, si adatta un modello preesistente per l'energia libera
        di un reticolo polimerico associativo al caso vitrimerico, esso dissociativo.
        Si sonda poi la validità del modello confrontandone le previsioni con i
        risultati di simulazioni di dinamica molecolare originali. Si dimostra che
        tali fusi vitrimero-polimerici possono subire separazione di fase anche
        in assenza di interazioni energetiche tra le loro componenti, e che il
        diagramma di fase del fuso è qualitativamente simile al suo analogo
        dissociativo. \DIFadd{Infine, si scopre che il numero critico per la
        separazione di fase di siti funzionali collegati ad altre catene
        vitrimeriche --- nel limite di catene isometriche e vitrimeri altamente
        funzionalizzati --- è di due siti a catena.}
	\end{abstract}
    \begin{description}\small
    \item[\textbf{Parole chiave}] vitrimeri, fusi polimerici, separazione di fase,
        smiscelamento entropico, reticoli covalenti adattabili associativi,
        reticoli polimerici dinamici, \DIFadd{legami bifunzionali}
    \end{description}
\end{otherlanguage}
\vfill~\clearpage

\section{Introduction}\label{sec:intro}

Fifteen years ago, Leibler and co-workers \autocite{montarnal2011} first stably
synthesised a new class of materials they dubbed \enquote{vitrimers}.
These materials have a wide variety of applications \autocite{zheng2021,%
denissen2016}, which include their use as a %
recyclable, self-healing alternative to thermosets; as a compatibiliser
providing reversible adhesion between immiscible materials; or, mixed with a
classical homopolymer, as an additive to improve thermomechanical properties
without loss of reprocessability \autocite{wang2021,torres2026}. They achieve this
remarkable versatility by forming polymer networks much like a thermoset, with
a fixed number of cross-links, but capable of reversibly reshuffling cross-link
sites to rearrange the network topology. Vitrimers are built from functional copolymers,
typically obtained from commercially available monomers with the periodic
addition of pendant functional groups (\ie, cross-link--capable sites).
A small cross-linking molecule is then added, forming covalent bonds
with the functional sites; being highly energetic, these bonds are rarely broken at
normal temperatures. However, they can exchange attachment points with other
cross-linkers, and do so often, as this reaction has fairly low energy of
activation \autocite{winne2019}. Sites not occupied by cross-linkers are
stoppered by some \enquote{stopper} moiety, which acts as
a place-holder to prevent empty sites from reacting haphazardly. The only reaction
present in the final material is thus the exchange of attachment points between
moieties (cross-linker--cross-linker and cross-linker--stopper).
In this way, a dynamic network is formed which
rearranges itself topologically yet always maintains a constant overall number
of cross-links.%

\begin{figure}[htb] %
	\centering
	\begin{subfigure}[b]{0.3\linewidth}
		\centering
		\includegraphics[width=\linewidth]{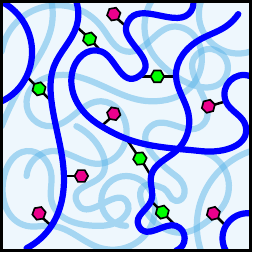}
		\caption[Cartoon of the melt]{%
            Cartoon of the melt, consisting of the polymer~(\tikz{
                \draw[as A, ultra thick, scale=0.3] plot coordinates {
                (0.2,0.4) (0.8,0.7) (1.4,0.3) (1.8,0.6) };
            }\kern.2em/\kern.08em\textcolor{A}{A}), the vitrimer backbone~(\tikz{
                \draw[as B, ultra thick, scale=0.3] plot coordinates {
                    (0.2,0.4) (0.8,0.7) (1.4,0.3) (1.8,0.6) };
            }\kern.2em/\,\textcolor{B}{B}), the cross-linker~(\tikz{
                \node[as C, scale=0.7] (loneC) at (1,0.5){};
                \draw[line CD] (0.7,0.5) to (loneC.corner 3);
                \draw[line CD] (1.3,0.5) to (loneC.corner 6);
            }\kern.2em/\,\textcolor{C}{C}) and the stopper moiety~(\tikz{
                \node[as D, scale=0.7] (loneD) at (1,0.5){};
    		      \draw[line CD] (0.7,0.5) to (loneD.corner 3);
            }\kern.2em/\,\textcolor{D}{D}).%
        }%
		\label[figure]{fig:melt}
	\end{subfigure}
	\hfill
	\begin{subfigure}[b]{0.6\linewidth}
		\centering
		\includegraphics[width=\linewidth]{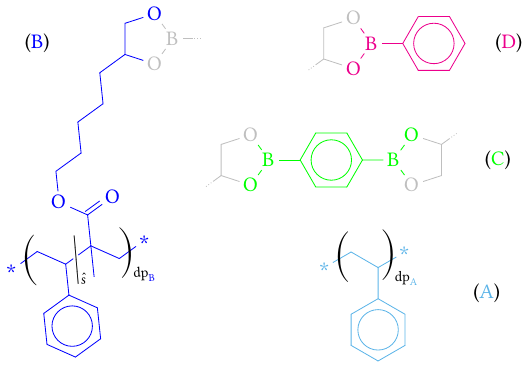}
		\caption[Polystyrene-based vitrimer melt]{Chemistry of a polystyrene
        vitrimer melt with dioxaborolane cross-linker and stopper.}
		\label{fig:PS}
	\end{subfigure}
	\caption[Anatomy of a melt]{Anatomy of a vitrimer--homopolymer melt.}
	\label{fig:anatomy}
\end{figure}

One application of particular interest involves dispersing vitrimers into a
classical homopolymer matrix. By carefully selecting the materials involved,
the resulting mixture can be made more thermomechanically resistant than a
traditional thermoplastic, at a fraction of the cost of a pure vitrimer.
Unlike a thermoset with permanent cross-links, this hybrid material is
reprocessable and self-healing, thanks to the cross-links' dynamic nature.
The anatomy of such a mixture --- in the specific case of a polystyrene vitrimer
with dioxaborolane moieties --- is illustrated in \cref{fig:anatomy}. Therein, the
cross-linker moiety~(C) attaches to functional sites on the vitrimer backbone~(B),
and the stopper moiety~(D) is simply a monofunctional version of the
cross-linker, filling up all remaining sites. The fourth component is the
matrix~(A), which we refer to interchangeably as polymer or homopolymer.

Studies of such melts are still scarce. However, preliminary evidence suggests
that such melts do not mix homogeneously. A study of poly(methyl methacrylate) in
an epoxy vitrimer matrix \autocite{han2018} finds clear micron-scale phase separation
when the mix is equal parts by weight. Analyses of pure vitrimers
\autocite{hayashi2025, ricarte2019} also detect traces of phase separation
at several scales; see \textcite{deheerkloots2023} for a detailed qualitative
discussion. Controlling micro-structure is important for the practical use of
vitrimers: some applications call for well-mixed melts; others rely on specific
modes of phase separation \autocite{joosten2024}.
Currently, achieving the correct micro-structure
relies on repeatedly iterating experiments and prototypes, a procedure both
costly and time-consuming. There is thus a need for a predictive approach to
phase separation, which can be achieved through an appropriate theoretical model.

Classical theories of adaptive polymer networks apply to \enquote{dissociative}
systems, where the functional sites detach and reattach without the intermediary
of cross-linker or stopper moieties. This phenomenon is governed by the energy
of association, which is the energy respectively lost or gained by
forming or breaking a bond. As such, dissociative networks create or destroy
cross-links at will, thereby freely altering their connectivity properties.
Since bonds can associate and dissociate, the actual number of bonds
\begin{enumerate*}[(i)]
    \item will fluctuate around the mean fraction for a material at equilibrium,
    \item may evolve with time when out of equilibrium, and
    \item may not be conserved during processes such as phase separation ---
          in other words, the total number of bonds may change.
\end{enumerate*}
The mean fraction of bonds associated at any given time is called the
\enquote{degree of conversion}~\((p)\), and can be predicted at equilibrium
by minimising the free energy of the dissociative system. The result
depends on the energy of association as well as on the
composition of the mixture; it also varies with temperature,
with hotter samples exhibiting fewer bonds (lower~\(p\)) on average.

In contrast, vitrimers --- or \enquote{associative} systems --- exist in the limit of high energy of
association, when it is almost never favourable to break a bond. This justifies the
assumption that, in vitrimers, the total degree of conversion is a conserved quantity:
in this approximation, cross-links can change attachment sites, but can never detach spontaneously.
A consequence is that vitrimer systems can be treated purely entropically:
since the number of bonds remains fixed, the total bonding energy remains constant;
hence, it is possible to conceive of an idealised, {entropic} vitrimer system
whose state depends only on the number of microstates generated by the presence of
bonds and different chain configurations. Although real vitrimer blends almost
certainly also involve energetic effects (\eg\ from interaction parameters),
it is instructive to investigate first the ideal, {entropic} case.
Therefore, we will restrict the scope of this paper to systems governed entirely
by entropic effects, without energetic interactions between any of the four
component moieties A--D (\cf~\cref{fig:anatomy}).

\pagebreak
Existing theories mostly describe dissociative networks; nonetheless,
similar systems have been extensively studied in the theoretical literature.
Approaches include mean-field theory \autocite{semenov1998,dobrynin2004}, a
patchy-particle model \autocite{smallenburg2013}, coherent-state field theory
\autocite{vigil2025}, cluster expansions \autocite{fantoni2011} and more; see
\textcite{tanaka2022} for a review. The mean-field theory due
to \textcite{semenov1998} and expanded by \textcite{dobrynin2004} is a popular
framework.
It has been used to study solutions of polyelectrolytes \autocite{chen2025},
blends of heteroassociative polymers \autocite{danielsen2023},
mixtures of polymers with compatible functional sites \autocite{prusty2018},
\DIFadd{single-chain nanoparticles with multiple sticker types \autocite{cappa2026},}
and has been to an extent compared with experiments \autocite{skorik2016}.
This approach has the advantage of producing a free-energy
function from counting arguments alone, without invoking heavy
mathematical artillery.
In this paper, we use this framework to arrive at a phase diagram of
vitrimer--polymer melts in the absence of enthalpic effects. We will find similar
qualitative predictions to those of dissociative
systems \autocite{semenov1998,dobrynin2004}, namely that macro-phase separation
occurs whenever the degree of conversion (resp.\ energy of association) is
sufficiently large.
In contrast with the dissociative case, however, there are compositions of melt
which never undergo separation (generally if the proportion of vitrimer is
sufficiently large or small). Furthermore, the separation we
predict is fully driven by the entropy of the resulting phases, meaning that it
is a purely statistical phenomenon.

\DIFadd{Phase separation driven by entropic effects is of course not new.  In the polymer literature previous works have considered phase separation induced by differences in chain stiffness, or short-chain branch content in polyolefins.  These  include lattice-based models \autocite{bawendi1986, foreman1997}, field-theoretic approaches \autocite{fredrickson1994}, and more recent simulations \autocite{kozuch2016, zhang2017a}.  However, these studies differ from ours in that the entropic effect is due to changes in the number of local packing configurations available when dissimilar polymers are mixed. Since these effects are local, they can be accounted for via an entropic contribution to the effective \(\chi\)~interaction parameter, as predicted by several of the above models. In contrast, the entropic effect at play in our study concerns changes in the translational entropy and network connectivity as composition is varied.  In this sense, the entropic effects in our work are more analogous to those previously identified in polymer--microemulsion networks \autocite{zilman2003}.}

Our work addresses the need for a predictive model of phase separation in
vitrimer--polymer melts by providing an initial, simplistic description using the
mean-field theory. Thus far, no systematic experimental campaign has been
conducted to capture regimes of stability in these melts. To validate our
description, we therefore turn to numerical methods.
Molecular dynamics simulations have been used
extensively to model polymer systems
\autocite{Kremer1990DynamicsSimulation,Everaers2020,Behbahani2025}; by adding
Monte-Carlo-based interactions between selected beads, they have successfully
been extended to model dynamic polymer networks \autocite{Cui2023LinearLimits,
Xia2023StructureVitrimers, Perego2022MicroscopicVitrimers}.
While Monte-Carlo simulations have been used extensively to study the
\emph{viscoelastic} behaviour of vitrimer melts, phase separation remains largely
unexplored territory. To our knowledge, this study constitutes the first
systematic comparison with simulations of the mean-field-type
theory introduced by \textcite{semenov1998}.

The goal of this paper is, therefore, to provide a combined theoretico-simulative approach
to predicting macroscopic phase separation in vitrimer--polymer melts, meant to
guide experimental exploration of these materials' design space.
The paper is structured as follows. In \cref{sec:theory}, we derive a mathematical
model for the free energy of a vitrimer--polymer melt devoid of energetic
interactions. \Cref{sec:simulation} then presents the hybrid Monte-Carlo\,/\,%
molecular-dynamics simulative approach adopted to validate the theory. In
\cref{sec:discussion}, we compare simulation results with the theoretically
predicted incidence of macroscopic phase separation, and discuss possible
extensions of the theory. Conclusions are drawn in \cref{sec:conclusion}.

\section{Free energy model} \label{sec:theory}

In this section, we derive the Helmholtz free energy for a vitrimer--polymer
melt. For the sake of exposition, \DIFadd{and for better comparison with the simulations carried out in \cref{sec:simulation},} we will only present a simplified derivation
where the small moieties --- cross-linker and stopper --- are \enquote{phantom}
objects which do not occupy any volume.
\DIFadd{We will discuss how this assumption can be lifted in
\cref{sec:realisticphasediagram}, with a complete derivation given in the
supplementary information (\cref{apx:fullderivation}).}
\DIFdel{(A complete derivation with this assumption lifted is given in the supplementary information, appendix~A.)}%
The resulting\DIFadd{, simplified} model assumes that the functionalised monomers on every vitrimer chain can
directly bond to other functionalised monomers, without the intermediary of a physical
cross-linker. We will use the language of previous works on dissociative
polymer networks and call these special monomers \enquote{stickers}, and
define a sticker as \enquote{bonded} if the equivalent functional site is
attached to a cross-linker, and \enquote{free} otherwise
(\cref{tab:terminology}).  Though confusing at first,
this change in language allows better alignment with existing literature, and
makes the link to simulations more straightforward.

\begin{table}[htb]
	\centering
	\begin{tabular}{ll}
		\toprule
		Chemical nomenclature & Phantom model \\ \midrule %
		Functional site & Sticker \\
		Site with cross-link & Bonded sticker \\
		Site with stopper & Free sticker \\ \bottomrule
	\end{tabular}
	\caption[Terminology conversion]{Conversion between terminologies.}
	\label{tab:terminology}
\end{table}

Therefore, the melt consists of two species: a homopolymer matrix, chemically inert,
and a vitrimeric network of the same chemical nature,
every chain of which has \(s\)~designated stickers.
We assume that stickers are uniformly distributed along the chain (\ie, located at
equal monomeric intervals), with half an interval at the extremities.
To emulate the effect of covalent bonds, we fix the total number of
sticker--sticker cross-links (\enquote{bonds}) to some global value.
Since the number of bonds is fixed, and the vitrimer and homopolymer
are assumed not to interact energetically, all microstates have the
same enthalpy, and the internal energy may be excluded from the
computation of the partition function.
\DIFadd{%
Even if a system phase separates, with different amounts of
cross-linking in each phase, the constant \emph{total} number of bonds (summed over both phases) 
ensures the total internal energy is constant. As bonds never break
without immediately reforming, their association and dissociation can
never contribute to the overall energy balance.
}%
The calculation thus involves
only a conformational or entropic term, counting the number of ways
to arrange vitrimer and polymer chains in the melt.
The predictions of this simplified model will be shown
at the end of \cref{sec:phasediagram},
and later compared to the results of simulations in \cref{sec:comparison}.

\subsection{Derivation}\label{sec:sr}

We follow in large part the derivation of \textcite{semenov1998}, rewritten in
terms of a lattice model. The major difference is that we do not approximate the
solvent's effect through its virial coefficients, given that our \enquote{solvent}
is actually a polymer matrix. Our basic theory, before the addition of
cross-links, is the classical Flory--Huggins theory of polymer melts.

To obtain the partition function, we count the number of ways of arranging the
vitrimer and polymer chains on a lattice, subject to the constraint of forming
the correct number of sticker--sticker bonds. We start by giving the number of
ways of arranging the chains in the absence of such constraints,
\(\Omega_\text{FH}\), and later correct it by a combinatorial factor.

Given a lattice of~\(n_0\) sites filled with \(n\)~vitrimer chains, each
occupying \(N\)~sites, and \(m\)~polymer chains, each occupying \(M\)~sites, the
number of ways of placing these chains on the lattice is given in the mean-field
approximation by \begin{equation}\label{eq:fhOmega}
	\Omega_\text{FH} \coloneq \frac{n_0!}{n_0^{n_0}} \frac{(n_0\zeta_N)^n}{n!} \frac{(n_0\zeta_M)^m}{m!} \,,
\end{equation}
where the \(\zeta_i\) represent the number of internal configurations available
to the chains on the lattice, uncorrected for excluded volume; hence the explicit
dependence on \(N\) and~\(M\) in  \(\zeta_N\) and~\(\zeta_M\).
\Cref{eq:fhOmega} is a standard result of the Flory--Huggins theory; its
derivation can be found in \textcite{flory1953} or, for convenience, in the
supplementary information (\cref{apx:fullderivation}).

If there are \(c\)~cross-links between stickers (one cross-link for two stickers),
the partition
function~\(\Omega_\text{FH}\) needs to be multiplied by the number of ways
of forming \(c\)~distinct sticker pairs, as well as by the probability that
the chains are in a configuration allowing that many bonds. Forming \(c\)
pairs out of \(s \times n\) stickers is equivalent to ordering the stickers in a list
defined such that the first sticker is paired with the second, the third with the
fourth, and so on. The remaining \(sn-2c\) stickers are unpaired. The following
diagram illustrates the situation, with each sticker represented by
an asterisk~\((*)\). %
\[\underbrace{(*,*)\, (*,*)\, \cdots \, (*,*)}_{\substack{2c\text{ stickers} \\[.2em] \text{ form }
c \text{ pairs}}} \ds\bigg|\ds
\underbrace{{*}\,{*}\,{*}\,{*}\,{*}\, \cdots
\, {*} \vphantom{()}}_{\substack{sn-2c\text{ stickers} \\[.2em] \text{remain unpaired}}} \]
There are~\((sn)!\) ways of arranging the stickers in this list, to be
corrected by dividing by~\(c!\)~(since the ordering of the pairs doesn't matter),
by~\((sn-2c)!\) (since the ordering of the unpaired~stickers doesn't matter),
and by~\(2^c\) (since the ordering of the stickers within a pair doesn't
matter).

Next we compute the probability that the chains be in a configuration allowing
\(c\)~bonds. A given bond is allowed if and only if that bond's second sticker
occupies one of the \(x\)~sites on the immediate coordination shell of the first
sticker. This happens with probability~\(x/n_0\). The event then
needs to recur \(c\)~times, so the probability becomes~\((x/n_0)^c\).
(The number of cross-link-compatible second sites~\(x\) depends on the specific
geometry of the chains, but is usually taken to be two less than the lattice
coordination number, accounting for monomers adjacent to the sticker.)

The full partition function~\(\Omega\) is thus the product of these two factors
with~\(\Omega_\text{FH}\), namely
\begin{equation}\label{eq:srOmega}
	\Omega = \Omega_\text{FH} \times \frac{(sn)!}{2^c \, c! \, (sn-2c)!}
	\times \left(\frac{x}{n_0}\right)^{\!c}\,.
\end{equation} To obtain an expression for the entropy, we apply Stirling's
formula \(z!\sim(z/\ec)^z\) and take the natural logarithm of~\(\Omega\).
We also wish to rewrite the expression in terms of (independent) intensive
variables: one is the volume fraction occupied by vitrimer chains, \(\phi
\coloneq Nn/n_0\); the other is a rescaled version of the volume fraction
occupied by bonded stickers: \(\psi \coloneq 2c/n_0 \times N/s\). We choose the
scaling factor \(N/s\) on this second
quantity so that \(\psi\) ranges between zero and~\(\phi\), and the degree of
conversion of stickers is \(p \coloneq 2c/sn = \psi/\phi\).
The volume fraction occupied by the polymer is --- by conservation of volume
--- not an independent variable, and indeed is equal to \(1-\phi = Mm/n_0\).
Performing these substitutions, and defining \(\mathfrak{H}(z)\coloneq z\log(z)\) for
compactness, the non-dimensional free energy per lattice site
\(f\coloneq -\log(\Omega)/n_0\) is given by \begin{equation}\label{eq:srf}
\frac Ns f \approx \frac1s \left[\mathfrak{H}(\phi) + \frac{N}{M} \mathfrak{H}(1-\phi)\right] %
 + \mathfrak{H}(\phi-\psi) + \frac12 \mathfrak{H}(\psi) - \mathfrak{H}(\phi)\,.
\end{equation}
Here and elsewhere in the text we have freely removed from the free energy both constants and terms linear
in the concentration variables~\((\phi,\psi)\), as they have
no effect on the resulting predictions (\cf~\cref{sec:phasediagram}).
It is worth emphasising that the expression \(Nf/s\) in~\cref{eq:srf} is
independent of any lattice parameters. Note also that the behaviour of the system does
not directly depend on the lengths of the component chains, but only on their ratio \(M/N\).

Heuristically, this expression for the free energy can be seen as the interplay
between two competing influences: the first term in parentheses ---
scaling as \(1/s\) --- is convex, encouraging stable behaviour,
whereas the rest of the expression pushes for instability.
As~\(s\) increases, the unstable influence becomes more prominent
and the region of instability grows, leaving only a small island of stability
near \((\phi, \psi)=(1,0)\).

A variant derivation, due to \textcite{dobrynin2004}, suggests that the
expression in \cref{eq:srf} takes into account the presence of cycles in the
network; the details of this, along with the full derivation for volume-occupying
cross-linker and stopper, are given in the supplementary information (\cref{apx:fullderivation}).

\subsection{Nearest-neighbour correction} \label{sec:nn}

\begin{figure}[htb]
    \centering
    \includegraphics{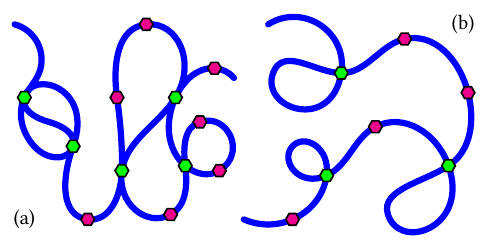}
    \caption[Nearest-neighbour bonds]{Generic same-chain linking (a) \vs\ nearest-neighbour only (b).}
    \label{fig:nearestneighbourbonds}
\end{figure}

The partition function provided in \cref{eq:srOmega} assumes that the second
sticker within each pair has equal probability of lying anywhere in the system
volume. Whilst good for stickers disconnected from each other (or only connected
through a long enough path), this assumption is deficient when stickers are
separated by short lengths of vitrimer backbone. In this case, the stiffness of
the chains affects sticker distributions. The effect is particularly salient when
the vitrimer is dilute in the matrix, and stickers mostly see stickers from the
same chain in their immediate vicinity.  %
Following both \textcite{semenov1998,dobrynin2004}, we model these local bonds
through their leading-order behaviour, when stickers pair with their nearest
neighbour along the chain (\cref{fig:nearestneighbourbonds}).
Longer-range bonds are polynomially suppressed by the
entropic cost of forming a loop in the polymer, with scaling \((\text{loop
length})^{-3/2}\).

First we calculate the contribution to the partition function due to
nearest-neighbour bonds. Let a fraction~\(q\) of the \(s\)~stickers on a given
vitrimer chain be involved in nearest-neighbour pairs. To calculate how many ways
this can happen,
imagine having a list of the \((s-qs)\)~free stickers and \(qs/2\)~cross-linked
sticker pairs on that chain, and asking the number of ways to rearrange these.
By a \enquote{stars-and-bars} argument \autocite[\eg][]{levin2021},
this is the same as taking \((s-qs)+qs/2\)~objects, and selecting \(qs/2\) to be pairs.
The number of ways of doing this is simply the binomial coefficient \((s - qs/2)!
/ [(qs/2)! \, (s-qs)!]\). However, forming \(qs/2\)~nearest-neighbour pairs
comes with an entropic cost, as the configurations that the two bonded stickers
can have relative to each other are reduced. This cost can be accounted for by
multiplying the partition function by the probability~\(\mathbb
P[{l}_{N/s}]\) of forming a loop of length~\({l}_{N/s}\),
repeating as many times as there are loops~(\(qs/2\)).
The product of these two expressions is the amount by which each vitrimer chain
modifies the partition function due to local bonds between nearest-neighbour
stickers.
For the full partition function, we multiply \cref{eq:srOmega} by
\(n\)~copies --- one for each vitrimer chain --- of this local-bond correction
term. The result is \begin{equation}\label{eq:nnOmega}
	\Omega = \Omega_\text{FH} \times \left(\frac{(s - qs/2)!}{(qs/2)! \, (s-qs)!}
	\mathbb P[{l}_{N/s}]^{\frac{qs}{2}} \right)^{\!n} \times
	\left(\frac{x}{n_0}\right)^{\!\tilde c} \frac{(\widetilde{sn})!}{2^{\tilde c}
	\, \tilde c! \, (\widetilde{sn}-2\tilde c)!} \,,
\end{equation} where \(\widetilde{sn} = sn-qsn\) and \(\tilde c = c - qsn/2\)
are the values reduced by the number of stickers (resp.\ cross-links) involved in local,
nearest-neighbour pairs. As before, \cref{eq:nnOmega} can be used to provide the
free energy, yielding \begin{equation}\label{eq:nnf}
\begin{split}
	\frac Ns f &= \frac1s \left[\mathfrak{H}(\phi)+\frac NM \mathfrak{H}(1-\phi)\right] +
	\mathfrak{H}(\phi-\psi) \\
	&\hspace{4em} + \frac12 \mathfrak{H}(\psi-q\phi) - \mathfrak{H}\!\left([1-\frac 12q]\,\phi\right) \\
	&\hspace{8em} + \frac12 \mathfrak{H}(q\phi) +  \frac{1}{2} q\phi \log\left(\frac{2}{k\ec}\right) \,.
\end{split}
\end{equation} The first line is identical to its equivalent in \cref{eq:srf},
the second line has an equivalent but now depends on~\(q\), and the
third line is entirely new.  We recover \cref{eq:srf} in the limit \(q\to0\),
namely when all cross-links are to other vitrimer chains.
(We have assumed so far that the cross-linker is two-sided: a similar expression
to \cref{eq:nnf} holds when the cross-linker is of arbitrary functionality; this
derivation is given in the supplementary information, \cref{apx:funrxlk}.)

In \cref{eq:nnf}, the  fraction~\(q\) of local cross-links is an unconstrained
variable obeying only the inequality \(q\leq p\). We therefore fix its value by
extremising the free energy over \(q\), %
namely by setting \(\partial{f} / \partial{q} = 0\).
The resulting optimality condition
\begin{equation}\label{eq:nnq}
	\frac{(2-q)\,q}{p-q} = \frac k \phi \,
\end{equation}  is controlled by a parameter \(k \coloneq \mathbb{P}[{l}_{N/s}]
\times 4N/(sx)\) measuring the propensity for local bonding.
\Cref{eq:nnq}
can be solved to give  \begin{equation}
q = \frac{k + 2\phi}{2\phi} \left(1 - \sqrt{1 - \frac{4kp\phi }{(k+2\phi)^2}}\right) \,.
\end{equation}

A good approximation is afforded by linearising~\(q\) with respect to~\(p\),
yielding \(q \approx kp/(k+2\phi)\). This approximation is excellent so long as
\(4 k p \phi /(k + 2\phi)^2 \ll 1\); it is at its worst when \(\phi=k/2\) and
\(p=1\), with maximal error \(3/2-\sqrt{2}\), or about \(9\,\%\). This error
reduces quickly with decreasing~\(p\), being just over \(5\,\%\) by \(p=0.8\),
and has
little impact on the majority of the explored phase space; for typical values of
\(p\) in practical applications, on the order of a few units per cent, the error
is less than one in five thousand.
This approximation greatly increased the speed in our numerical computation
by facilitating evaluation of the free energy and its derivatives.

We see from the linearisation that the simpler expression in \cref{eq:srf} is most
accurate when \(k\to0\), which occurs in the limit of greater inter-sticker
spacing since \(k\sim (N/s)^{-1/2}\) (see next section).
Note also that as the solution becomes more dilute, we see \(q\) approach~\(p\):
in other words, local bonds dominate the cross-linking behaviour,
and percolation becomes increasingly difficult.

\subsection{Local bonding propensity}

The parameter~\(k\) controlling the local bonding propensity %
remains a free parameter of this theory. Some intuition for it can be
afforded by rephrasing it in terms of material properties of the vitrimer.
In this section, we provide an interpretation through a simplified scaling
argument then, for reference, perform the calculation exactly for ideal
(Gaussian) vitrimer chains.
The latter, exact result is not essential to the discussion.

Consider a segment of chain between two stickers, consisting of~%
\({l}_{N/s}\) Kuhn segments, each of length~\(b\). To find~\(k \coloneq
\mathbb{P}[{l}_{N/s}] \times 4N/(sx)\), we need the probability~%
\(\mathbb{P}[{l}_{N/s}]\) of the segment's two ends lying within a
bond volume~(\(v_\text{b}\)) of one another. This probability is
proportional to the bond volume~\(v_\text{b}\) when \(v_\text{b}\)
is small, since the underlying segment-end distribution is approximately
constant over small regions. The maximum value that the bond volume can
take --- at which the probability should be unity --- scales as the cube
of the chain segment's root-mean-square end-to-end radius~\(R_\text{s}
\sim b \sqrt{{l}_{N/s}}\), which characterises the typical distance
by which segment ends are likely to be separated. Thence the scaling
\(\mathbb{P}[{l}_{N/s}] \sim v_\text{b} / R_\text{s}^3\). Replacing the
bond volume with its lattice-based equivalent \(v_\text{b} =
x \kern0.05em v_0\), we obtain the scaling law~\(k \sim
v_\text{s} / R_\text{s}^3\), where \(v_\text{s} = v_0 N/s\) is the
volume taken up by the inter-sticker chain segment, or equivalently the
volume of vitrimer chain per sticker. Interpreting \(R_\text{s}^3\) as the
volume spanned by the chain section between two stickers, we find that
\(1/k \sim R_\text{s}^3/v_\text{s}\) is (roughly) the typical number of
stickers within a volume~\(R_\text{s}^3\) for a pure vitrimer system ---
\ie, in the absence of a homopolymer matrix.
Its square~\(1/k^2\) is also known as the \enquote{invariant degree of
polymerisation} \autocite{qin2024} for that inter-sticker segment.

The packing length~\(P\) of a polymer chain corresponds to the length
scale on which a chain fully fills its pervaded volume, and is defined as
the ratio of the chain's displaced volume to its root-mean-square end-to-end
radius \autocite{bobbili2021}. Using the chain segment between adjacent
stickers, we have \(P \sim v_\text{s}/R_\text{s}^2\). The consequent scaling
\begin{equation}\label{eq:kpacking}
    k \sim P / {R_\text{s}}
\end{equation} gives a further intuitive interpretation of~\(k\) as the
reciprocal inter-sticker distance, measured in units of the packing length.
This can be understood as encoding the fact that stickers within the same
packing blob are easy to cross-link without losing much entropy; the more
blobs they are separated by, however, the more costly (and therefore rare)
cross-links become, and \(k\to0\).

For reference, we now compute~\(k\) theoretically for an ideal chain
\autocites[§\,2.3]{doi1988}[§\,40.10]{grosberg1994}. Recall that the
Green's function~\(\mathcal G_{l}(\vec{r})\) for an isolated phantom chain
of \({l}\)~effective bonds of length~\(b\) is given by the expression
\begin{equation}
	\mathcal G_{l}(\vec r) = \left(\frac{2\pc}{3} {l} b^2\right)^{-3/2}
    \exp\left(-\frac{3 \left\Vert\vec{r}\right\Vert^2}{2 {l} b^2}\right)
\end{equation} where \(\vec r\) is the end-to-end vector of the chain.
Requiring that the chain ends lie within a volume~\(v_\text{b}\) of one
another is then equivalent to integrating over the locality
operator~\(v_\text{b} \, \delta\), where~\(\delta\) is the Dirac delta.
Thus \begin{equation}
\mathbb{P}[{l} ] = \int_{\mathbb{R}^3} \mathrm{d}^3\vec r \  v_\text{b}\,
\delta(\vec r) \, \mathcal{G}_{l}(\vec r) = \frac{v_\text{b}}{b^3}
\left(\frac{2\pc}{3} {l} \right)^{-3/2} \,.
\end{equation}
The last step is then to express the inter-sticker distance in units of
the effective bond length~\(b\). This is simply the number of Kuhn segments
between successive stickers, obtained by dividing the volume \(v_\text{s}
= v_0 N/s\) of an inter-sticker chain segment by the volume \(v_\text{K}\)
of a Kuhn segment, giving \({l}_{N/s} = N/s \times v_0 / v_\text{K}\).
Substituting all of these into \(k \coloneq \mathbb{P}[{l}_{N/s}] \times
4N/(sx)\) and again replacing \(v_\text{b} = x \kern0.05em v_0\) gives
\begin{equation}\label{eq:kgauss}
    k =  \frac{3\sqrt{6}}{\pc^{3/2}} \sqrt{\frac s N} \left(
	\frac{v_\text{K}}{\sqrt[3]{v_0}\,b^2}\right)^{3/2} \,,
\end{equation}
with numerical prefactor about~\(1.34\).
In particular, note the scaling \(k \sim \sqrt{s/N}\), %
implicit in~\cref{eq:kpacking}.

\subsection{Phase diagram}\label{sec:phasediagram}

From the free energy, we can obtain the phase diagram, displaying the system's
behaviour for the entire phase space \((\phi,p) \in {[0,1]}^2\), with
\(p=\psi / \phi\). We first describe how the various features of the diagram
are computed, before presenting a typical phase diagram at the end of the section.

\subsubsection*{Methodology}

The spinodal curve is the transition from local stability to local instability,
and occurs when the free energy goes from concave in both directions to convex in
at least one; we find this curve by numerically searching for the zero contour of the
Hessian determinant of the free energy \(f=f(\phi,\psi)\). Namely, we solve
\begin{equation}\label{eq:spinodal}
	f_{,\phi\phi}f_{,\psi\psi} -(f_{,\phi\psi})^2 = 0 \,,
\end{equation} where the subscript comma denotes differentiation with respect to
subsequent variables.
The binodal, marking the transition from true stability (global convexity) to a
metastable state (locally convex but globally concave) is less straightforward,
but is guaranteed to intersect the spinodal in at least one point,
\DIFadd{called the \enquote{plait point}. The plait point is a particular type of
critical point in liquid--liquid systems corresponding to the location where
binodal points --- the extremities of tie-lines, \ie, coexisting phases ---
become identical and coalesce
\autocites(\qqv){shcherbakova2023}[§\,2.12]{tompa1956}.}
Our methodology begins by obtaining this intersection
by simultaneously solving the aforesaid spinodal
condition --- \cref{eq:spinodal} --- alongside an equation
specific to the plait point:
\begin{equation}\label{eq:plait}
	f_{,\psi\psi\psi} f_{,\phi\psi} f_{,\phi\phi}
	- 3 (f_{,\phi\psi})^2 f_{,\phi\psi\psi}
	+ 3 f_{,\phi\psi} f_{,\psi\psi} f_{,\phi\phi\psi}
	- (f_{,\psi\psi})^2 f_{,\phi\phi\phi}
	= 0  \,.
\end{equation}
\Cref{eq:plait} corresponds to the statement that the plait point is an
inflection point on the spinodal curve; it can be derived from the condition
\((\vec t \cdot\nabla)^3 f=0\), where \(\vec t \propto [-f_{,\phi\psi} \ds
f_{,\phi\phi}]^\top\) is any tangent to the spinodal and \(\nabla \coloneq
[\partial_\phi \ds \partial_\psi]^\top\) is the \DIFdel{differential operator}\DIFadd{gradient} on the
phase space of conserved variables.
Then, the binodal is found by imposing (and numerically solving for) equality of the chemical
potentials \(\mu_\phi \coloneq f_{,\phi}, \ \mu_\psi \coloneq f_{,\psi}\) and
osmotic pressure \(\Pi \coloneq \phi \mu_\phi + \psi \mu_\psi - f\) across both
phases, according to the common-tangent construction. Straight-line segments
joining such solutions are known as tie-lines, and conservation of the phase
parameters \((\phi,\psi)\) implies that any unstable state lies on the tie-line
into whose extremities it will separate. Therefore, we use a stepping algorithm
to generate the tie-lines by guessing an unstable state and finding the
tie-line passing through it, starting at the plait point and always moving
perpendicularly to the last tie-line found. The set of tie-line extremities
forms the binodal.
\DIFadd{(This is one of several possible approaches:
\textcite{archer2024} outline a similar algorithm, based on sweeping
through chemical potentials instead of unstable states;
other options are the iterative method, as in \textcite{qian2022},
or the method of dominance analysis \autocite{qian2024}.)
Our} algorithm is prone to pitfalls such as local optima and
three-phase separation; to avoid these, we used a more robust but less accurate
method to first visualise the region of instability. Namely, we computed the
lower convex envelope of the free energy surface, since it is flat in \(i\)
directions when coexistence is between \((i+1)\) phases, \(i\in\{1,2\}\). We
are grateful to Peiwen Ren \autocite[see][]{ren2023} for his advice and example
code performing this computation.
Finally, the mean-field percolation transition is obtained by solving an
appropriate Flory--Stockmayer condition
\autocites[eq.\,17]{dobrynin2004}[fig.\,18a]{santra2021}\relax\space
\begin{equation}\label{eq:percolation}
	p-q = \frac{1}{(1-q)s-1}
\end{equation}
for the sticker conversion~\(p\) as a function of the vitrimer
fraction \(\phi\), with \(q=q(p,\phi)\) from~\cref{eq:nnq}.
\DIFadd{Under this condition, we say a system is in a \enquote{sol} state when vitrimer chains form finite clusters, and in a \enquote{gel} (or percolated) state as soon as there is one (infinite) cluster spanning the whole system.
\Cref{eq:percolation} is derived by assuming clusters are tree-like, with no cycles, and computing the expected number of new chains at each generation:
the criterion for percolation is when this value is greater than
unity \autocite[§\,6.4]{rubinstein2003}.}

\subsubsection*{Phase diagram}

\begin{figure}[hbt]
	\centering
	\includegraphics%
		{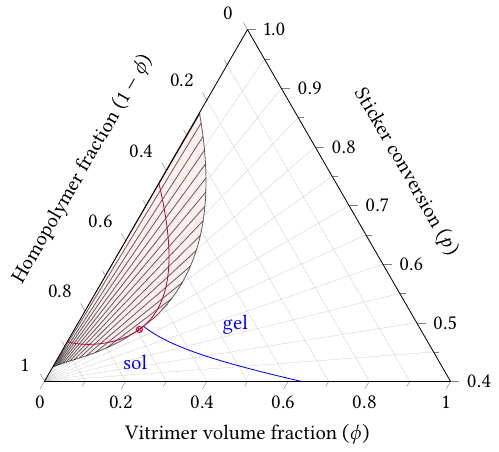}
	\caption[Theoretical phase diagram]{Phase diagram obtained from \cref{eq:nnf}
	with \(M=N=20\), \(s=5\) and \(k=0.48\), showing the gelation transition
	\DIFadd{(blue), spinodal curve (red), binodal curve (grey), tie-lines (grey
	segments) and plait point (symbol \(\otimes\))}.
	\emph{Data are plotted on \enquote{pinched axes}
	as the phase space is naturally triangular. On these axes, lines of constant
	\(p\) are rays emanating from the left-hand corner, whereas lines of constant
	\(\phi\) connect a value on the vitrimer axis with its complement on the
	homopolymer axis. Note that we show only a subspace \(p\geq0.4\).
	}}
	\label{fig:phasediagram}
\end{figure}

\Cref{fig:phasediagram} shows the resulting phase diagram, for parameter values
chosen to match the simulations performed in \cref{sec:simulation}. \DIFadd{The diagram displays both a percolation (sol--gel) transition and a miscibility dome.  Any system with volume fraction and sticker conversion lying outside the miscibility dome will be single phase: either a sol below the gelation transition or a gel above it. Within the miscibility dome, the system will phase separate.}
We note that instability tends to set in as cross-links are added to the system,
globally raising~\(p\). In the upper part of the phase diagram, \DIFadd{within the miscibility dome, }the two phases resulting from phase separation \DIFadd{are indicated by the extremities of the
tie-lines that intersect the system-averaged volume fraction,~\(\phi\), and sticker conversion,~\(p\). On increasing~\(p\) away from the plait point, these tie-lines} increasingly share the same local \(p\)~value after splitting.
\DIFadd{For this choice of parameters, most trial phases separate into both sol
and gel phases, since the percolation transition passes very near to the plait point.
The separation into sol and gel appears to be a common feature of these
polymer--vitrimer systems, as will later be discussed for~\(s\gg1\)
(\cref{sec:scaling}). However, the gelation line does not exactly meet the plait point, and for these parameters we predict a minority of systems close to the plait point that phase separate into two sol phases.}

Qualitatively, our phase diagram is similar to those of
\textcite{semenov1998,dobrynin2004}. Heuristically, the globally conserved
sticker conversion (\(p\)) in associative networks can be compared to the
attractive volume of a bond (\(\lambda\), which depends exponentially on the
energy of association) in dissociative networks: both control the density of the
resulting network, and for the purposes of visualisation can be plotted as the
ordinate of the phase diagram against the vitrimer concentration variable on the
abscissa. Doing so, we observe the same topological features, namely two-phase
regions for high \(p\) (resp.~\(\lambda\)) often --- but not always ---
separating between a sol and a gel phase.  Of course, the difference in control
parameters between associative and dissociative polymer networks means that the
correspondence is imperfect, and predictions can only ever be compared
qualitatively.

\section{Simulation methodology}\label{sec:simulation}

To validate the theoretical framework proposed in \cref{sec:theory}, molecular 
dynamics simulations were performed to investigate phase 
separation phenomena in purely entropic vitrimer--homopolymer blends. \DIFadd{Here, we present
simulations of the simplified model without the inclusion of explicit cross-linker molecules.}
We employ the widely adopted 
coarse-grained Kremer--Grest bead-spring model 
\autocite{Kremer1990DynamicsSimulation} utilising fully flexible chains without angular potentials. A system of 2000 chains, each of length $N=20$ beads, 
were initialised in a cubic periodic box with size $L \gg 5 \,  R_\text{g} $ (where 
$ R_\text{g} $ is the mean radius of gyration), such that the monomer number 
density was $\rho=0.85 \, \sigma^{-3}$. 

The vitrimer volume fraction $\phi$ was
varied in the interval $\phi \in [0.1, 0.9]$, in steps of~$0.1$.
Vitrimer chains are distinguished from
homopolymers by the inclusion of evenly distributed monofunctional associative
groups (stickers) along the backbone. \DIFadd{The only difference between stickers and regular beads is
the stickers' ability to form crosslinks with a maximum of one other sticker. Regarding their mass
and physical interactions, the two types of monomer are identical.}
\DIFdel{Specifically,}Each vitrimer chain contains
five stickers, which are distributed along the twenty-bead backbone at positions
2, 6, 10, 14, and 18. A functionalisation of five stickers per chain ensures the
system resides well above the percolation threshold (given sufficiently large
sticker conversion is achieved), thereby minimising possible topological defects. Furthermore, it
allows reasonable computation times.

\subsection{Equilibrium molecular dynamics}

Excluded volume interactions between non-bonded beads are modelled by the purely
repulsive Weeks--Chandler--Andersen (WCA) potential
\autocite{Weeks1971RoleLiquids}, a Lennard-Jones (LJ) potential truncated and
shifted at $r_\text{c} = 2^{1/6} \, \sigma$:
\begin{equation}\label{eq:LJ_potential}
    U_{\text{LJ}}(r) =
    \begin{cases*}\
        4 \epsilon \left[ {\left(\frac{\sigma}{r}\right)}^{12} - {\left(\frac{\sigma}{r}\right)}^{6}\right] + \epsilon & for \(r \leq r_\text{c}\) \\
        \hfill 0 \hfill & for \(r > r_\text{c}\)
    \end{cases*}
\end{equation}
where $r$ denotes the inter-particle distance, while~$\epsilon$ and~$\sigma$
represent the fundamental energy and length scales respectively. The simulations
employ standard reduced units with bead mass~\(m_0\) and Boltzmann constant~\(k_\text{B}\).
Consequently, time is expressed in units of $\tau \coloneq \sigma \, {(m_0/\epsilon)}^{1/2}$.

Connectivity between adjacent backbone monomers, as well as the reversible
cross-links formed between stickers, is governed by a finitely extensible
non-linear elastic (FENE) potential \autocite{Kremer1990DynamicsSimulation}. This
potential superposes the attractive entropic spring with the purely repulsive WCA
component to prevent \enquote{bond crossing} (\ie, bonds passing through each
other), thus respecting topological constraints:
\begin{equation}\label{eq:FENE_potential}
    U_{\text{FENE}}(r) =
    \begin{cases*}\
        -\frac{1}{2}KR_0^2 \log \left[ 1-{\left(\frac{r}{R_0}\right)}^2 \right] + U_{\text{LJ}}(r) &  for  \(r < R_0\) \\
        \hfill \infty \hfill &  for  \(r \geq R_0\)
    \end{cases*}
\end{equation}
where $K=30 \, \epsilon/\sigma^2$ is the spring constant and $R_0=1.5 \, \sigma$ represents the maximum bond length.

\DIFadd{The above interaction potentials mean that the simulations are not purely entropic, and include some energetic (enthalpic) contributions.  However, we omit any excess energies of mixing between the homopolymer and vitrimer species
(\ie~$\chi=0$) by employing the same non-bonding interactions between beads of the two species and by using the same bonding potential for both species.
The \enquote{phantom} nature of the cross-links ensures that there is also no excess energy of
mixing between the cross-links and anything else.  Hence, entropy is the dominant factor driving any phase separation observed.}

Prior to the activation of reversible cross-linking, the polymer melt underwent a multi-stage equilibration protocol maintained at a reduced temperature $T^*=1 \, \epsilon/k_\text{B}$ using a Langevin thermostat ($\tau_{\text{damp}} = 0.1 \, \tau$). To eliminate initial steric overlaps and thermally equilibrate the system, non-bonded interactions were initially governed by a soft repulsive potential. This relaxation phase proceeded for a total of $2 \times 10^8$~time steps, beginning with an integration step of $\chg t = 0.001 \, \tau$ which was subsequently increased to $\chg t = 0.005 \, \tau$ to accelerate relaxation. Following the removal of high-energy overlaps, the standard Lennard-Jones potential of \cref{eq:LJ_potential} was restored, and the system was equilibrated for a further $10^8$~time steps with $\chg t = 0.001 \, \tau$.

To closely control sticker conversion, we developed a controlled cross-linking procedure. Every 10~time steps, unreacted stickers were allowed to form a maximum of 1 bond with another sticker using the \verb.fix bond/create. command available in LAMMPS \autocite{LAMMPS}. The capture radius was set to $R_\text{min}=1.4 \, \sigma$ to restrict bonding to the first coordination shell, and prevent high energy bonds from forming near \(R_0\). Secondly, the probability of creating a bond between two candidate stickers was set to \(P_{\text{bond}}=0.0001\) to allow the gradual build-up of cross-linking density. Sticker conversion~($p$) was monitored using an external Python script; as soon as the desired cross-linking density was achieved, bond creation was turned off and the mixture was equilibrated in a canonical (NVT) ensemble with time step $\chg t = 0.01 \, \tau$ for a total of $10^8$~time steps. The bond potential of the formed cross-links is described by \cref{eq:FENE_potential}. At this stage, all cross-links between stickers were permanent in nature.

Following equilibration and cross-linking, dynamic bond exchange reactions were activated. These measurement runs were conducted in a canonical ensemble, where temperature was maintained at $T^*=1 \, \epsilon/k_\text{B}$ using a Langevin thermostat ($\tau_{\text{damp}} = 0.5 \, \tau$). Newton's equations of motion were integrated with a time step of $\chg t = 0.01 \, \tau$ for a total duration of $10^8$~time steps. Every $10^3$~time steps, bond exchange reactions were allowed to occur using the algorithm described in \cref{subsec:BERs}. Initiating data collection immediately upon the activation of bond exchanges allowed us to monitor the phase separation kinetics in real time, capturing the system's evolution from the precise moment topological reconfigurations began.

\subsection{Bond exchange reactions}\label{subsec:BERs}

Vitrimers are classified as associative covalently adaptive networks (asso-CANs)
\autocite{denissen2016}. A defining characteristic of this class of material is the strict
conservation of cross-link number, maintained through an exchange mechanism where bond cleavage is
immediately followed by bond formation \autocite{ricarte2019}. To capture this topology-altering
behaviour in the melt state, hybrid Monte-Carlo\,/\,molecular dynamics (MC/MD) algorithms have been
used extensively \autocite{Perego2020VolumetricStudy, Perego2021EffectVitrimers,
Perego2022MicroscopicVitrimers, Cui2023LinearLimits, Xia2023StructureVitrimers}. Building upon the
framework established by \textcite{Cui2023LinearLimits}, we extended their MC/MD algorithm to
explicitly incorporate exchange reactions involving non-bonded, \enquote{free} stickers.

\begin{figure}[htb]
    \centering
    \subcaptionsetup{aboveskip=-0.5em}
    \begin{subfigure}[b]{0.49\linewidth}
        \centering
        \includegraphics[width=\linewidth]{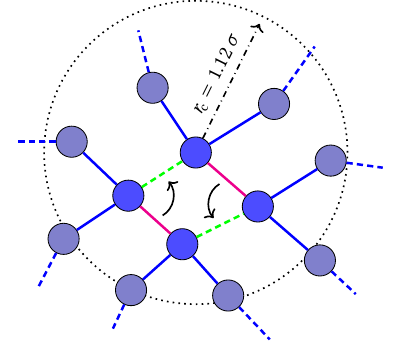}
        \caption{Bond swap}
        \label{subfig:BER1}
    \end{subfigure}
    \hfill
    \begin{subfigure}[b]{0.49\linewidth}
        \centering
        \includegraphics[width=\linewidth]{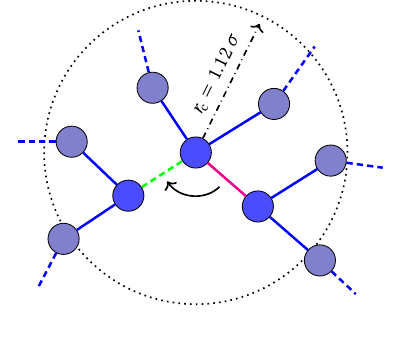}
        \caption{Bond shift}
        \label{subfig:BER2}
    \end{subfigure}
    \caption[Types of bond exchange reaction]{Bond exchange
    reactions (BER).}
    \label{fig:BERs}
\end{figure}

Bond exchange can occur through two distinct reaction pathways. A \enquote{bond swap} (\cref{subfig:BER1}) involves the metathesis between two existing cross-linked sticker pairs, whereas a \enquote{bond shift} (\cref{subfig:BER2}) is facilitated by the presence of a free sticker in the immediate vicinity of a bonded pair. Crucially, both reaction types are subject to a strict proximity criterion, requiring all participating stickers to be within a defined cut-off radius $r_\text{c} = 1.12 \, \sigma$. This cut-off value corresponds to the minimum of the WCA potential (\(2^{1/6} \, \sigma\)), representing the direct contact distance between non-bonded stickers. This strict spatial constraint for bond exchange attempts serves two purposes: first, it mimics the steric requirements of an associative reaction mechanism, ensuring the new bonding partner is in the immediate coordination shell; second, it ensures that newly formed cross-links are created near their equilibrium length (\(\approx 0.96 \, \sigma\)), minimising high-energy stretched states and eliminating the possibility of bond crossing upon formation of new cross-links. Consequently, the local spatial distribution of stickers dictates the eligibility of specific sites for exchange events.

For each possible proposed move, the change in potential energy is calculated as:
\begin{equation}
    \Chg U = \sum U_{\text{FENE}}^{\text{new}}(r) - \sum U_{\text{FENE}}^{\text{old}}(r)
\end{equation}
where \(\sum U_{\text{FENE}}^{\text{new}}(r)\) and \(\sum U_{\text{FENE}}^{\text{old}}(r)\) are the cumulative FENE potentials of the newly proposed (post-exchange) and the original (pre-exchange) bonds, respectively. For both reaction pathways depicted in \cref{fig:BERs}, two distinct new configurations are possible. Candidates were screened by energy minimisation: namely, we retained only the configuration with the lowest \(\Chg U\).

The acceptance of the proposed bond exchange reaction (BER) is then governed
by the standard Metropolis criterion \autocite{Metropolis1953}. Moves
resulting in a net reduction of potential energy ($\Chg U \leq 0$) are
accepted unconditionally, whereas energetically unfavourable ($\Chg U >0$)
BERs are only accepted with an exchange probability
\begin{equation}\label{eq:Pacc}
    P_{\text{acc}} = \exp \left( - \frac{\Chg U}{k_\text{B} T^*} \right) \,.
\end{equation}

To eliminate the possibility of bond exchange reversals within a given
integration cycle, stickers that have participated in a successful exchange
are excluded from further bond exchange attempts during that cycle. This
means that each sticker is limited to a maximum of one bond topological
change per MC step.

Strict detailed balance is violated in the algorithm due to the sequential
processing of bond exchange candidates, as the exclusion of reacted stickers
from subsequent moves within the same time step introduces a non-Markovian
dependency on the random iteration order. Nonetheless, the bond exchange
reaction algorithm does satisfy the weaker criterion of global balance,
which is the necessary and sufficient requirement for Monte-Carlo simulations
\autocite{Manousiouthakis1999}. On the other hand, when screening the two
proposed configurations by energy minimisation, a deterministic choice is
introduced which technically violates the balances.

Since the difference in potential energy between the two candidates is
generally a multiplicative factor, the higher energy candidate would have
an acceptance probability several orders of magnitude smaller due to the
exponential suppression in \cref{eq:Pacc}. Therefore, the error introduced
by this choice is negligible, and it has the benefit of allowing a significant
speed-up in the Monte Carlo algorithm. Thus, this preference towards lower
energy transitions identifies dominant reaction pathways, effectively
filtering out statistically negligible high-energy transitions. This is
applicable for the temperatures generally used in Kremer--Grest simulations
(\(T^* \lesssim 2 \, \epsilon/k_\text{B}\)).

\subsection{Inferring stability}\label{sec:inferring}

In simulations, it is possible to observe the development of phase separation over time. Starting from an equilibrated, permanently cross-linked network, BERs are turned on and the measurement run starts. The state of the system is inferred at regular time intervals by recording atom coordinates and measuring the distribution and local concentration of beads belonging to vitrimeric chains (both ordinary beads and stickers).

Local properties are calculated by discretising the simulation box into $7^3 = 343$ equal-volume subdomains. To mitigate bias arising from arbitrary boundary placement, the discretisation grid was shifted incrementally along all three axes in steps of one-tenth the subdomain side length. These choices --- dividing each side of the simulation box into seven equally sized segments and performing ten grid shifts in each dimension --- were guided by two considerations. First, both numbers are coprime, ensuring that no two sampled configurations are identical. Second, splitting the simulation box into seven equal segments along each axis yields a subdomain length of approximately~\(5 \, \sigma\), which is large enough to contain a sufficient number of beads to average over, while remaining small enough to produce observable differences between subdomains. Taking into consideration the periodic boundary conditions, this generates an ensemble of $343\,000$ subdomains. Here, we describe three distinct methods of determining stability from these discretised subdomains.

\subsubsection*{Distribution plots}

The simplest way to infer the stability of a system is by looking at the number
distributions of vitrimeric beads in the (equally sized) subdomains. Stable
systems exhibit the expected Gaussian-like distribution, whereas for unstable systems,
a split in the distribution is observed, indicative of two distinct phases
forming within the melt.

\begin{figure}[htb]
	\centering
	\includegraphics[width=3.33in]{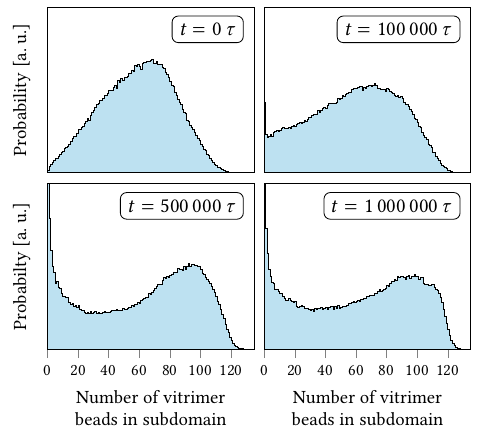}
	\caption[Atom count distribution]{Atom count distribution of vitrimeric beads
		at four different times during simulation (\(\phi=0.5\),
		\(p=0.95\)). \em The entire simulation lasted a total of \(t=10^6 \, \tau\), or \(10^8\) time steps.}
	\label{fig:ACD}
\end{figure}

The development of these distributions over time (\cref{fig:ACD}) can give information regarding the rate at which phase separation is developed and when equilibrium is reached. Results reveal that equilibration is invariably reached by the half-way point of the entire simulation (\(10^8\)~time steps), with only slight, stochastic deviations observed afterwards.

Distribution plots prove useful to determine phase separation
for well separated systems, where both phases occupy reasonable fractions
of the simulation box. For such systems, the distribution plots show clear
bimodal distributions. However, for \enquote{edge} cases where bimodality is not
apparent, more advanced tools are needed to quantitatively determine
phase separation. For this purpose, two alternative methods to infer
stability are presented, based on the spatial correlation function and on
density variance, respectively.

\subsubsection*{Spatial correlation}

\begin{figure}[htb]
	\centering
	\includegraphics[width=3.33in]{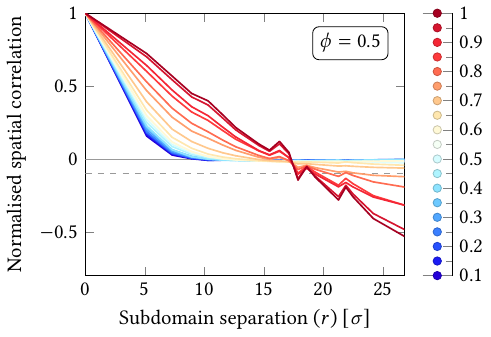}
	\caption[Spatial correlation]{Spatial correlation, for \(\phi=0.5\) and
		different sticker conversions \(p\), as a function of the distance
		between subdomain centres. \em The cut-off point for
		determining stability is indicated by the horizontal dashed line.
		The same plots for other values of	\(\phi\) can be found in the
		supplementary information (\cref{apx:plots}).}
	\label{fig:spatial_corr}
\end{figure}

The stability of the systems were primarily determined from the spatial
correlation function. First, we calculate the local composition \(\phi(i)\) of
the \(i\)\textsuperscript{th}~subdomain, defined as the proportion of beads in
the subdomain that belong to vitrimer chains. This is then correlated with
that of all other subdomains as a function of their centre-to-centre Euclidean distance~\(r\)
according to the formula \(\sum_{ij}\left[\phi(i) - \phi\right]\left[\phi(j) - \phi\right]\), where \(\phi\) is the global average composition and the sum is taken over all subdomain pairs \((i,j)\) separated by a fixed distance~\(r\).
An example of this correlation is shown in \cref{fig:spatial_corr} for \(\phi=0.5\).

Formally, for a macroscopic system, phase separation would correspond to an infinite correlation length appearing in the system. Non-phase-separated systems would then have a spatial correlation function that decays to zero over a finite distance. Unfortunately our simulations are necessarily finite, with a simulation box made even smaller by its periodic boundaries. This means that: \begin{enumerate*}[(i)]
    \item the largest correlation length available corresponds to about half the diagonal of the simulation box, \ie~about \(30\,\sigma\);
    \item a phase-separated system will have strong positive correlations in composition up to around half the simulation box size, and negative correlations at large distances because of the overall mass conservation in the finite box; but
    \item such negative correlations are also observed as we approach the region of instability in parameter space, because the diverging correlation length quickly approaches the simulation box size.
\end{enumerate*}
As a result, it is difficult to distinguish the exact point at which phase separation occurs in a finite simulation box: for intermediate cases it may be impossible to decide between a system with long correlations near the phase-separated state, and a system that is actually phase separated. Hence, plotting the spatial distribution function for different values of~\(p\) gives a continuum of curves rather than a discontinuous change at phase separation. As a practical choice, we decided to take an anticorrelation of~\(-0.1\) at large distances as a heuristic criterion for phase separation. It should be noted that this value is most likely size-dependent and may require adjustment for different box volumes. As can be seen from \cref{fig:spatial_corr}, this criterion also corresponds roughly to the point where the zero of the spatial correlation function approaches that found for the clearly phase separated systems (at larger~\(p\)).

\subsubsection*{Density variance}\label{sec:density_var}

To support the conclusions drawn from the spatial correlation on the stability of a system, we also look at the change in density variance over time. It is defined as \(\sigma_\rho(t)^2 = \langle \rho_i(t)^2 \rangle - \langle \rho_i(t) \rangle^2 \), where \( \rho_i (t)\) is the vitrimer bead density for each subdomain~\(i\) at time~\(t\), and \(\langle \ \cdot\ \rangle\) is the average over subdomains;
\DIFadd{this definition implies \(\sigma_\rho(t)^2 \propto \sum_i [\phi_i(t)-\phi]^2\), where \(\phi\) is the global average composition and the sum is over subdomains~\(i\).}
When normalised by the density variance at \(t=0\) (the moment before which only permanent cross-links existed between stickers), stable and unstable configurations show opposing behaviour.

\begin{figure}[htb]
	\centering
	\includegraphics[width=3.33in]{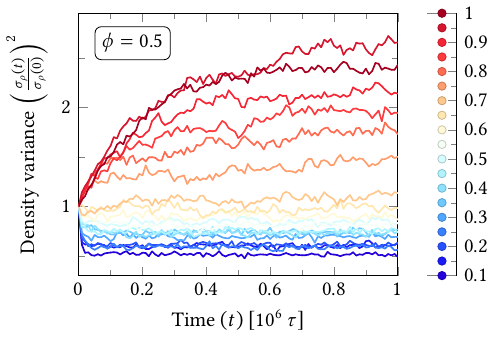}
	\caption[Density variance]{Density variance as a function of time for
	\(\phi=0.5\) at different sticker conversions \(p\).
	\em The same plots for other values of \(\phi\) can be found in the
	supplementary information (\cref{apx:plots}).}
    \label{fig:density_var}
\end{figure}

\Cref{fig:density_var} shows the density variance as a function of time for systems with \(\phi=0.5\). The different lines correspond to different sticker conversions (\(p\)). Systems that phase separate show an increase in their normalised density variance over time, until a stable plateau is reached halfway through the simulation. An increase in density variance is indicative of melt heterogeneity (phase separation). Conversely, stable blends show a decrease in variance, synonymous with increased homogeneity and therefore stability. We classify systems into three categories according to the value of their density variance at large times: less than \(1\), stable; between \(1\) and \(1.5\), ambiguous; above \(1.5\), unstable. We note, as we did for spatial correlation, that these thresholds may change depending on the volume of the simulation box.

When looking at \cref{fig:density_var} a question arises: why does the density variance for stable systems decrease relative to the starting state? This phenomenon can be attributed to the increased mobility induced by BERs. By turning on BERs, the system is able to equilibrate further than it was previously capable to with permanent cross-links, and the increased mobility of beads permits a more homogeneous state.
This argument supports the idea that an increase in density variance can be attributed to phase separation.

\subsection{Percolation transition}\label{sec:gelation}

It is also important to conceptually validate the Flory--Stockmayer prediction of
the gelation line. Gelation is defined as the first point at which an infinitely
large network forms, spanning the entirety of the sample. In MD simulations, this
is
the moment at which the probability of finding a cluster spanning the entire
simulation box becomes non-zero by forming an \enquote{infinite} network through
the periodic boundaries.

We follow the method developed by \textcite{Livraghi2021} to determine the
percolation threshold from MD simulations and compare it to the prediction given
by \cref{eq:percolation}. Their estimate threshold is defined by the reduced
molecular weight (RMW) and point of inflection of the largest-mass cluster. The
RMW is defined as the molecular weight average of all clusters in the system
except the largest one. The largest cluster's size is a more indirect measurement
of gelation, where the moment at which the mass of the largest cluster starts to
significantly outweigh the other clusters is used as the gelation point. Here, we
use both methods to get the most reliable estimate of the critical sticker
conversion (\(p\)) needed to achieve gelation. The use of both methods is
justified considering that the values of \(p\) were varied in steps of \(0.05\),
which may be too low of a resolution for an exact match between theory and
simulations.

\begin{figure}[htb]
	\centering
	\includegraphics%
		{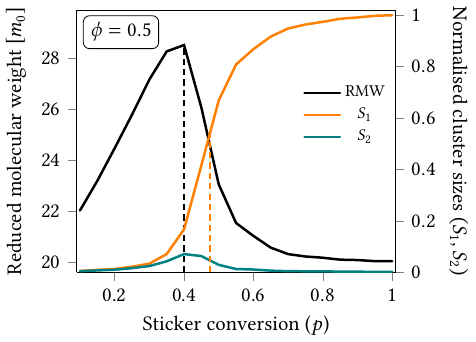}
	\caption[Percolation transition]{Percolation transition measurements for
	\(\phi=0.5\). \em Black curve: reduced molecular weight (RMW); orange
	curve: normalised largest cluster size \((S_1)\); teal curve: normalised
	second largest cluster size \((S_2)\). Dashed lines indicate the
	maximum of the RMW and the inflection point of \(S_1\). The same plots for
	other values of \(\phi\) can be found in the supplementary information
	(\cref{apx:plots}).}
	\label{fig:percolation}
\end{figure}

\Cref{fig:percolation} demonstrates the two methods described above for the
estimation of the gel point at a vitrimer fraction \(\phi=0.5\). The result
obtained from the maximum of the RMW curve predicts a lower critical value for
the onset of percolation, \(p=0.4\). On the other hand, the inflection point of
the largest cluster size estimates a higher onset (\(p=0.475\)). Using these two
methods of determining the sol--gel transition, we can provide a confidence
interval for percolation in the simulated systems. This result can then be
compared to the prediction given by the mean-field theory in
\cref{eq:percolation}.

\section{Results and discussion}\label{sec:discussion}

\subsection{Comparison between theory and simulations}\label{sec:comparison}

\begin{figure}[htb]
	\centering
	\includegraphics%
		{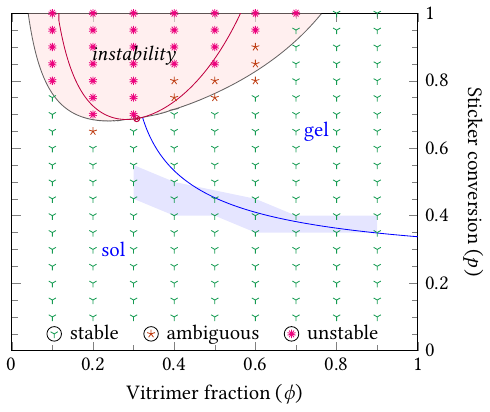}
	\caption[Comparison of theory and simulations]{
		Phase diagram from theory (lines) and from simulations (symbols).
		\emph{
			Three classes of stability are shown: stable, clearly
			phase separated, and ambiguous or borderline.
			The free-energy parameters are \(M=N=20\), \(s=5\), and
			\(k=0.48\). This figure shows the same theoretical data as
			\cref{fig:phasediagram}, plotted on square axes to maximise
            visibility of the simulation points for small~\(\phi\).
            The percolation transition seen	in simulations
            (\cref{sec:gelation}) appears as a pale blue band.%
		}
	}
	\label{fig:comparison}
\end{figure}

We are finally ready to compare the theoretical phase-diagram predictions with
the results of simulations. We initialise the free energy with parameter values
dictated by the simulations: choosing a discretisation such that every bead
fills exactly one lattice site, the parameters \(M = N = 20\) and \(s = 5\) are
straightforward from \cref{sec:simulation}. The free parameter~\(k\) was
estimated by visually fitting to the simulation results, and found to be
somewhere around one half: \(k= 0.48\pm 0.08\). (This motivated
the earlier choice of values in \cref{fig:phasediagram}.) For better visibility of the
simulated data, we refrain from using \enquote{pinched axes} in favour of a
more traditional square plot, giving \cref{fig:comparison}.
Note that on square axes, tie-lines are \emph{not} straight line segments,
and thus are not shown. On the same plot, we show the spinodal curve
(red line, inside the region of instability),
\DIFadd{the plait point (symbol \(\otimes\))}
and the mean-field percolation transition (blue line).

The results of simulations are split into three classes --- stable, clearly
phase separated, and borderline --- and shown on \cref{fig:comparison} using
symbols. The confidence interval for the percolation transition, obtained
using the methods in \cref{sec:gelation}, is indicated by a pale blue region.
The region forms a band around the theoretical prediction for \(\phi\geq0.5\),
when the mean-field theory is expected to be most accurate; the agreement
is worse for smaller values of~\(\phi\), when same-chain bonds become important
and the mean-field assumptions weaken.

Looking at the distribution of simulation results, we see that all systems which
displayed clear phase separation in the simulations lie within the theoretical
region of instability. There is some uncertainty near the boundary, due to the
gradual onset of separation in simulations as the \(p\)-value increases (as
discussed in \cref{sec:inferring}). Nonetheless, we find very good agreement
between theory and simulations at an appropriately chosen value for the free
parameter~\(k\).

\subsection{Consistency of the local-bonding parameter}\label{sec:validatek}

How reasonable is this fitted value \(k \approx 0.48 %
\)? We can check consistency by calculating it in different ways. One is to use
the result for Gaussian chains, \cref{eq:kgauss}, applied to a lattice based on
the MC/MD beads. Here, the bond length is around the minimum of the FENE potential
\((b\approx0.96\,\sigma)\); every bead is a statistical segment, so the Kuhn volume
is that of a bead \((v_\text{K} = \frac\pc6 \, \sigma^3)\); and every
lattice site corresponds to one bead \((v_0 = v_\text{K})\).
Using these values in \cref{eq:kgauss} gives \(k\approx0.39\),
in agreement with the fitted uncertainty \(0.48\pm0.08\).

\begin{figure}[htb]
	\centering
	\includegraphics%
		{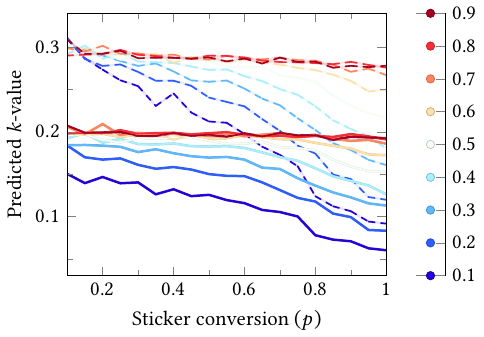}
	\caption[Bonding propensity]{Value of the left-hand side of \cref{eq:nnq} as
	a function of \(p\) for all values of \(\phi\). \em Solid lines indicate
	calculations involving only nearest-neighbour cross-links, dashed lines
	correspond to calculations involving all same-chain cross-links.}
	\label{fig:k_value}
\end{figure}

Whilst the %
accuracy of this agreement is to some extent coincidental --- for example, if
instead of using as unit volume the size of a bead \((v_0=v_\text{K})\) we had used
the inverse number density of monomers in the simulation box \((v_0=1/\rho\approx
1.18\,\sigma^3)\), the estimate would have risen to \(k\approx0.88\) --- it remains
heartening to find a result of the correct order of magnitude. This can be
compared with another way to verify consistency, namely measuring the values
of \(p\) and~\(q\) directly in the simulations and using them to solve
for~\(k\) in \cref{eq:nnq}. Limiting this analysis to systems
which unequivocally do \emph{not} undergo phase separation, we obtain
\(k\approx0.20\) from the fraction of cross-links which are between nearest-neighbour
stickers on the same chain. Note however that this is expected to be an underestimate:
in the simulated systems, same-chain bonds are not limited to being between
nearest-neighbour stickers, and can span the whole chain; nearest-neighbour bonds
are thus less likely than they might be in a nearest-neighbour-only system. To
compensate for this, we can imagine a different definition of~\(q\) in simulations:
instead of using the fraction of bonds which are between \emph{nearest-neighbour}
stickers, we can measure the fraction of bonds which are between \emph{any two
stickers on the same chain}. Using this alternative definition for~\(q\)
in \cref{eq:nnq} gives \(k\approx 0.30\) on average.
(\Cref{fig:k_value} shows the detail of the calculation for both \(q\)~variants,
plotting the equivalent \(k\)~value for each simulated sticker conversion \(p\) and
vitrimer volume fraction \(\phi\).) For stable systems, the
predicted values of \(k\) remain approximately constant, and the value starts to
drop as soon as instability sets in, since the single-phase percolation condition
ceases to apply. As with the previous method, we find a value of the correct order
of magnitude, though the estimate is now further from the fitted value.

A third method of controlling consistency is to return to the definition
\(k \coloneq \mathbb{P}[{l}_{N/s}] \times 4 N/ (sx)\). Assuming beads in the
simulation are to a good approximation hexagonally packed, each sticker
has roughly \(x = 12 - 2=10\) free sites adjacent to it. The natural
discretisation --- one bead per lattice volume --- makes \(N/s\) the
number of bead diameters between successive stickers. The only missing
quantity is then~\(\mathbb{P}[{l}_{N/s}]\), the probability that two
nearest-neighbour stickers are within a bond length of each other. This quantity can be obtained from the simulations by observing a system sure not to separate
--- \eg, \(\phi=0.9\) and \(p=0.5\) --- and counting, amongst all
nearest-neighbour sticker pairs, which proportion are less than a
distance \({%
                            2\times 0.96\,\sigma}\) apart. (In fact, to
avoid interference from the cross-linking algorithms, we look only at
beads on \emph{homopolymer} chains spaced four bond lengths apart.)
We find a probability \(\mathbb{P}[{l}_{4}] \approx 0.3\), yielding a
value \(k = 0.48\).

\subsection{Phase diagram for a realistic blend}\label{sec:realisticphasediagram}

Comparison with simulations suggests good reason to be confident in the applicability of the theoretical model. In this section, we provide a phase diagram for a more realistic choice of parameters, inspired by a polystyrene-based vitrimer chemistry.

\subsubsection*{Steric cross-linker and stopper}

Thus far, we have approximated cross-linker and stopper moieties as
\enquote{phantom} objects, occupying no volume. Depending on the size of these
two moieties, however, the approximation may have limited applicability. To relax
this assumption, we split the vitrimer network into three types of component: the
chain backbone~(B), the cross-linker~(C) and the stopper~(D).  Each component ---
or, for simplicity, \enquote{species} --- is considered to occupy a
volume~\(N_i\) in discretisation units: the index~\(i\) runs over the four
species in our system, the remaining species (A) being the homopolymer chains.
This gives four volume fractions~\(\phi_i\) of which to keep track.

Repeating the argument in \cref{sec:theory} for a four-species system gives an intensive free energy  \begin{equation}\label{eq:nnfnophantom}
	\begin{split}
		f &=
		- \mathfrak{H}\!\left(\frac{s\phi_\B}{N_\B} - \frac qp \frac{\phi_\C}{N_\C}\right)
		+ \sum_i \frac{\mathfrak{H}(\phi_i)}{N_i}\\
		&\hspace{4em}+ \left[\mathfrak{H}(1-q/p) + \mathfrak{H}(q/p)
        + \frac qp \log\left(\frac{2N_\B}{\ec s k}\right)\right] \frac{\phi_\C}{N_\C}  \,,
	\end{split}
\end{equation}
with \(q \approx kp / (k+2\phi_\B)\), where the four volume fractions \(\phi_i\)
are related by the stoichiometric condition \(s\phi_\B/N_\B = 2\phi_\C / N_\C +
\phi_\D / N_\D\) and the --- obvious --- new volume balance, \(\sum_i \phi_i = 1\).
The degree of conversion is also updated, to \(p = 2N_\B \phi_\C / (s N_\C \phi_\B)\).
In the interest of brevity, the derivation of these results is relegated to the
supplementary information (\cref{apx:nn}).

The consistency of \cref{eq:nnfnophantom} with the expressions in \cref{sec:theory,sec:simulation} can be seen by
\begin{enumerate*}[(i)]
    \item relabelling \(N_\A = M\) and \(N_\B = N\),
    \item  setting \(\phi_\C = ({sN_\C}/{2N_\B})\ \psi\) and \(\phi_\D = ({s N_\D}/{N_\B})\ (\phi-\psi)\) and
    \item taking the volume balance \(\sum_i \phi_i=1\) in the limit of small cross-linker and stopper size \((sN_\C\sim sN_\D \ll N_\B)\) to claim that \(\phi_\B = \phi = 1-\phi_\A\).
\end{enumerate*}
It can be verified that the result of this procedure is the simplified system discussed up to now.
\DIFadd{%
    This consistency implies that the simplified model, wherein the
    cross-linker and stopper are \enquote{phantom} objects, %
    correctly accounts for the mobility of real, physical cross-linkers.
    In the full calculation, the cross-linkers' translational entropy is reduced by attaching the cross-linkers to stickers; the simpler calculation instead includes a combinatorial entropy of cross-linking, but the resulting contribution to the free energy is the same.
    The collapse of the two free energies --- \cref{eq:nnf,eq:nnfnophantom}
    --- in the limit of small cross-linker and stopper thus supports the
    intuition that the size of the cross-linker and stopper moieties has no
    impact on phase separation at zeroth order.
    In other words, the phase diagram varies continuously as the cross-linker size goes to zero.
    }

\Cref{fig:PS} showed the chemical structure of a polystyrene vitrimer--homopolymer
melt, categorised into our four species A, B, C and~D
(based on dioxaborolane functional groups).
The cutting-point between B and C/D is chosen to be the point where the chemical
bonds break during a cross-link exchange interaction. For such a system, what
values should be used for the parameters \(N_i\), \(i \in \{\text{A,\ldots,D}\}\,\)?
Recall that the on-lattice parameters \(N_i\) are defined as the number of
lattice sites occupied by a single representative of a given species. As such,
they are equal to the volume occupied by that representative divided by the unit
lattice-site volume: \(N_i = v_i/v_0\). Therefore, it suffices to estimate
the volume of one representative from each species, since the free energy is
invariant under rescaling of the elementary lattice volume~\(v_0\).

Given degrees of polymerisation \(\degreepolym \A\) and \(\degreepolym \B\) and
vitrimer functionality~\(s\), the volume occupied by each species can be
approximated by its mass. Counting atoms in \cref{fig:PS} gives \begin{equation}
		\begin{aligned}
				N_\A &= \degreepolym \A &\quad N_\B &= \degreepolym \B
				+\frac{47}{52} s \\
				N_\C &= 5/4 &\quad N_\D &= 1
			\end{aligned}
\end{equation} where the parameters \(N_i\) are generated by scaling every mass
by the same factor of \(104\,\text{Da}\).

\subsubsection*{Longer chains}

The degrees of polymerisation used for simulations --- twenty beads,
five stickers --- are constrained by the numerical capacity of the
computers at our disposal. In real applications, we might expect chains
to have monomer counts somewhere in the thousands, and function sites
numbering hundreds. We thus take our new-found, fully steric expressions
and apply them to a system with chains orders of magnitude longer.

As a plausible example, we consider both homopolymer and vitrimer to have degree of
polymerisation \(\degreepolym i = 1500\) and assume \(s = 200\) stickers per
vitrimer chain (corresponding to a molecular weight of \(175\,\text{kDa}\) for
the vitrimer and \(156\,\text{kDa}\) for the polymer).
For the unknown parameter~\(k\), we recall the
scaling law in \cref{eq:kgauss} and arbitrarily set \(k = (s/N_\B)^{1/2}
\approx0.34\). This is consistent --- recall \cref{eq:kpacking} --- with
values for the packing length \((P\approx 4\,\text{\AA})\) and root-mean-square
end-to-end distance \((R_\text{s}\approx 20\,\text{\AA})\) experimentally
measured for polystyrene \autocite{fetters1999,unidad2015}.

\begin{figure}[bht]
	\centering
	\includegraphics%
		{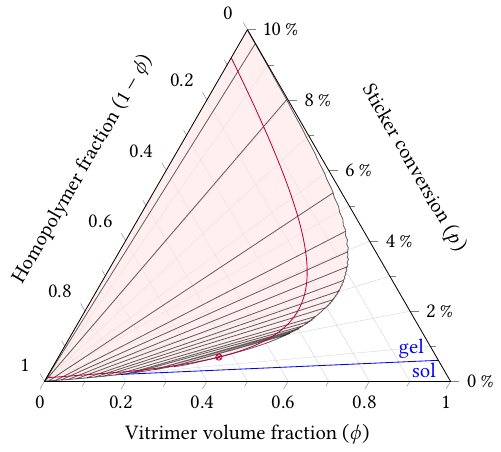}
	\caption[Phase diagram for polystyrene melt]{Phase diagram for a hypothetical
	poly\-sty\-rene-based vitrimer melt \DIFadd{with di\-oxa\-boro\-lane-based
	cross-linker and stopper moieties, showing the gelation transition (blue),
	spinodal curve (red), binodal curve (grey), tie-lines (grey segments) and
	plait point (symbol \(\otimes\))}.
	\emph{Parameter values are \(\degreepolym \A
	= \degreepolym \B = 1500\), \(s=200\) and \(k \approx
	0.34\). Since the \(p\)-axis starts at zero, horizontal sections of this plot
	correspond to slices of constant~\(\psi\).}}
	\label{fig:realphasediagram}
\end{figure}

\Cref{fig:realphasediagram} shows the resulting phase
diagram. (Due to lack of convergence of the numerical method in the
immediate neighbourhood of the plait point, the binodal in that region is
interpolated between the plait points and the first tie-lines identified.)
The most striking feature is the onset of instability, which now occurs
at much lower values of \(p\).
\DIFadd{In the next section (\cref{sec:scaling}), we show that this is consistent with scaling arguments for the simplified model of \cref{sec:theory} when~\(s\gg1\), which further suggests that, at zeroth order, the thermodynamics do not depend on the size of the cross-linker moiety.}

\DIFdel{
This last statement can be made more precise by solving for the plait point in
the simplest available model when chains do not bond to themselves ($k=0$).
This is equivalent to solving eqs.~(12) and (13) for the free energy
given in eq.~(3), with $N=M$. Doing so in a computer algebra system
provides an exact solution for the plait point, which can be asymptotically
expanded to first order in~$1/s$, yielding
$    \phi = \frac12 - \frac3{2s} + \bigO\left(s^{-2}\right)
    \ds\text{and}\ds
    \psi = \frac1s + \bigO\left(s^{-2}\right)
$
$
    \implies p = \frac2s + \bigO\left(s^{-2}\right) \,.
$
Whilst the correct estimate for the plait point is heavily dependent on the value
of~$k$, this equation gives a rule of thumb for the onset of instability:
the larger the number of stickers on the vitrimer chains, the smaller the critical
$p$~value for instability, roughly in reciprocal relation. We emphasise this
particular result, as it is helpful in guiding experimental exploration of the
design space of vitrimer--polymer blends.

The scaling $p\sim1/s$ can even be used to provide a
similar rule of thumb for the case with nearest-neighbour bonds $(k>0)$, through a
bootstrap-like argument. Repeating the preceding analysis with the corrected
free energy of eq.~(5), but setting $p=p_0/s$ and having the computer
algebra system take eqs.~(12) and (13) in the limit $s\to\infty$
gives the solution~$p_0=2+2k$, or equivalently $p \approx \frac2s \, (1+k)$
to first order.
Here, the average spacing of stickers along the backbone makes an appearance
through the dependence on~$k$, which by eqs.~(8) and (11)  scales as
$k\sim P/R_\textnormal{s}\sim \sqrt{s/\degreepolym\B}$. Increasing the density
of stickers along the vitrimer chain whilst keeping the functionality constant
is thus expected to displace the onset of instability towards higher degrees
of conversion.
}

\subsection{Asymptotics of highly functionalised chains}\label{sec:scaling}

\DIFadd{%
In the interest of quantifying the features of \cref{fig:realphasediagram}, we look for the position of the plait point and percolation transition asymptotically for large numbers of sticker sites per chain (\ie, as~\(s\to\infty\)). Their placement relative to each other will allow us to determine the generic appearance of the phase diagram in this limit. As the full model with steric cross-linker and stopper is analytically intractable, we return to the simpler model of~\cref{eq:nnf} for the asymptotic analysis.}

\DIFadd{%
    Finding the asymptotic behaviour of the plait point as \(s\to\infty\) requires simultaneously solving \cref{eq:spinodal,eq:plait} to first order in~\(1/s\). In the supplementary information, \cref{apx:pp-asymptotics}, we find this gives
    \begin{align}
    	\phi = \frac1{1+\sqrt{N/M}} - \frac{3\sqrt{N/M}}{2s} + \bigO\left(s^{-2}\right) 
    	\label{eq:pp-scaling-phi} \,,
    	\\[.5em]
    	\psi = \frac1{2s}\left(2+k+k\sqrt{N/M}\right) + \bigO\left(s^{-2}\right) 
    	\label{eq:pp-scaling-psi} \,.
    \end{align}
    From \cref{eq:pp-scaling-phi,eq:pp-scaling-psi}, we can compute the number of stickers per chain cross-linked to other chains at the plait point. This is the number of occupied sticker sites, \(ps\), reduced by the number of sites cross-linked to the same chain, \(qs\). To leading order in~\(1/s\), this gives
    \begin{equation}\label{eq:pp-scaling}
    	(p-q) \, s = 1 + \sqrt{\frac NM} + \bigO\left(\frac1s\right) \,.
    \end{equation} 
    In other words, for chains of comparable length \((N \approx M)\), a rule of thumb for the transition into immiscibility is that two stickers per chain be cross-linked to other chains. Comparing this to the rule of thumb for the percolation transition --- which is one sticker per chain --- we see that stability is only possible for networks which are loosely cross-linked, though it is possible to gel the system without phase separation as indicated in \cref{fig:realphasediagram}.
}

\DIFadd{%
When \(s\)~becomes large, \cref{eq:percolation} for the percolation transition becomes \((p-q)s\approx1\) to leading order, since \(q\leq p\in\bigO(1/s)\). Note the similarity to \cref{eq:pp-scaling}. Solving this explicitly gives \begin{equation}\label{eq:percolation-scaling}
	p \approx \frac1s \left(1 + \frac{k}{2\phi}\right)
\end{equation} at percolation. This is below the plait point 
but hits the region of instability fairly rapidly to the left of the plait point, since~\cref{eq:percolation-scaling} diverges as~\(\phi\to0\). Thus we see that, depending on the initial conditions, phase separation can either be into two gel phases or can span the percolation transition, giving both a sol and a gel phase. For large values of~\(s\), there is never separation into two sol phases.
}

\DIFadd{%
The vast majority of the phase space, where \(p\)~is large, separates into sol
and gel. In particular, as \(p\to1\), tie-lines become arbitrarily close to being
between \((\phi,\psi)=(0,0)\) and \((1,1)\), since the free energy
in~\cref{eq:nnf} becomes concave for \(s\to\infty\) and \(\psi\to\phi\).
Therefore, highly cross-linked systems typically separate into (almost) pure polymer and vitrimer phases.
In applications where the vitrimer is used as an additive to strengthen
the polymer without losing recyclability, this means that adding more covalent
bonds to the network will only cause it to separate more completely from the
thermoplastic matrix.
}

\subsection{Limitations of the model}

Certainly the primary limitation of the model presented is its pure-entropic
nature. In a real system, configurations have enthalpy: due to the interaction
between atomic potentials; due to compressibility of the sample; due to the
effect of temperature; --- none of this is accounted for in our model. Of these,
the energetic interactions of adjacent atoms are a particularly important %
omission. In the simplest case of a melt of identical backbone chemistries, as the
proportion of monomers which are functions (\(s/\degreepolym\B\)) grows small, we
expect the effect of energetic interactions to scale with the number of contacts
between homopolymer and vitrimer backbone on one side, and cross-linker
and stopper moieties on the other. Using the interaction parameter~\(\chi\) of
Flory--Huggins solution theory, this scale is roughly \(\chi s/N_\B\).
The energetic term will affect the position of the binodal line when it is
comparable to the stable, Flory--Huggins term of the free energy, which scales as
\(\max\{N_\A^{-1},N_\B^{-1}\}\). For equal chain lengths \(N_\A=N_\B\), this
suggests that the relevant parameter for energetic effects in our networks is
\(\chi s\), which is much less than the relevant parameter in homopolymeric melts
\(\chi N_\B\). We thus expect energetic effects to be suppressed in vitrimeric
melts.

\DIFadd{Like others before it, this model assumes that functional sites are evenly distributed along the vitrimer chain. In practice, an ideal co-polymerisation will lead to a random distribution of functional sites, and non-idealities such as variations in reactivity or co-monomer drift can lead to blockiness in site distribution or uneven distribution of sites between chains. This will inevitably affect the physics of the system.}

Another limitation is the model's treatment of the weakly concentrated regime (\(\phi\ll1\)).
In the presence of an energetic interaction with the solvent/matrix, previous literature
has suggested that the vitrimer chains would form globular structures
\autocite{semenov1998,dobrynin2004}. We believe this mechanism to be absent in
the entropic formalism presented in this paper. Nonetheless, a significant aspect
of the entropic case has been neglected in our treatment: namely, the
higher-order \enquote{harmonics} of same-chain cross-linking, which drive
chain collapse in real physical systems \autocite{berda2010}. We neglected
these under the assumption that they are suppressed by their polynomial entropic
cost, their contribution to the partition function scaling as \((\text{loop
length})^{-3/2}\). We know from \cref{eq:nnOmega} that the number of possible
nearest-neighbour-only configurations on \(s\)~sites with \(c\)~cross-links is
\((s-c)!/[c!\,(s-2c)!]\), and from \cref{eq:srOmega} that the total number of
configurations on \(s\)~sites with \(c\)~cross-links is
\(s!/[2^c\,c!\,(s-2c)!]\). This means that the proportion of configurations which
are nearest-neighbour only is the ratio of the first to the second, or
\begin{equation}
	\frac{2^c \, (s-c)!}{s!} \approx
	\exp\left[-c\log\left(\frac s2\right)+\bigO\left(\frac cs\right)\right] \,.
\end{equation} This quantity vanishes factorially as \(s\) grows large, largely
overtaking the suppression factor associated with non-nearest-neighbour
cross-links. This leads us to suggest that as the number of functions on a given
vitrimer chains grows, the cross-linking behaviour of isolated vitrimer chains is
not dominated by that between nearest-neighbour stickers, but exhibits cross-links
between stickers of arbitrary separation along the backbone.
This combinatorial effect will be compounded by the fact that a sort of
\enquote{economy of scale} is at play: once the vitrimer chain is already
collapsed, longer-range cross-links are no longer so entropically expensive as
they would originally have been. The topic of isolated vitrimer chains is still
an open question for research to tackle.

Finally, we wish to highlight that the mean-field methods used in this article
suffer the drawback that they are not spatially resolved: phases are treated
as homogeneous throughout. This means that such a theory cannot be used to make
meaningful statements about separation into micro-phases; yet if --- \eg\ ---
the cross-linker is strongly repelled by the vitrimer backbone, we expect the
formation of spatially segregated clusters of cross-linked sites.
An accurate treatment of such clusters may require the use of a fully-fledged
field theory \autocite[such as][]{vigil2025}, with the added complexity that such theories entail.

\section{Conclusion}\label{sec:conclusion}

In this study, we have derived a mean-field model for the free energy of an
associative covalent polymer network%
. From the free energy, quantitative predictions can be made for a wide range
of melt compositions and cross-link fractions by using appropriate numerical
methods.

The model's validity is supported by the agreement of molecular dynamics
simulations with the phase diagram predicted by the theory. These simulations
emulated the dynamics of vitrimer--polymer blends
by combining a Kremer--Grest bead-spring model with
a bespoke Monte-Carlo bond-exchange algorithm. Instability and percolation were
detected by comparing features in the distribution and clustering of vitrimer
beads within discrete subdomains of the simulation box; for instability, the
variance in bead density and the spatial correlation of the subdomains'
deviation from mean composition proved particularly useful in classifying edge
cases. Since the vitrimer and homopolymer analogues were made chemically
identical, any instability observed was solely a result of the entropic effect
of cross-links between stickers.

We predict phase separation for a range of parameter values, even
despite the exclusion of energetic effects. Stated differently: blends of
otherwise non-interacting polymeric materials can become immiscible merely
through the addition of a cross-linking mechanism.
This prediction, though consistent with previous work on dissociative systems
\autocite{semenov1998,dobrynin2004}, nonetheless has a novel aspect: only in
\emph{associative} systems is it possible to wholly disregard energetic
mechanisms. The phenomena described herein are entirely entropic, and serve as a
warning that affinity at a chemical level is in no way a guarantee that a
polymer--vitrimer blend will remain homogeneously distributed.

\DIFadd{Finally, we emphasise that phase separation sets in at low degrees of cross-linking --- as few as two sites per chain when the vitrimer and polymer are equal in nature and length, not much more than the amount required for gelation --- constituting an important physical limit to the use of vitrimers as a strength-enhancing additive.}

\nocite{rottger2017,ge2025}

\vfill
\[*\ds*\ds*\]
\vfill

\section*{Supplementary information}

Several additional sections are provided in the supplementary
information (at the end of this document). These
are: \begin{itemize}
	\item[(\ref{apx:fullderivation})] a full derivation of the free energy in the
	case of volume-filling cross-linker and stopper;
	\item[(\ref{apx:chi})] further discussion of the possible effects of energetic
	interactions in dissociative polymer networks;
	\item[(\ref{apx:funrxlk})] a derivation of the free energy (in the
	\enquote{phantom} approximation) in the case of arbitrary cross-linker
	functionality;
	\item[(\ref{apx:plots})] plots of the spatial correlation, density
	variance and percolation transition measurement for all values
	of~\(\phi\) (except \(\phi=0.5\), corresponding to
	\cref{fig:spatial_corr,fig:density_var,fig:percolation} in the text); and
	\item[(\ref{apx:pp-asymptotics})] \DIFadd{the calculation of the plait point asymptotically as \(s\to\infty\), justifying \cref{eq:pp-scaling-phi,eq:pp-scaling-psi,eq:pp-scaling} in the body of the text}.
\end{itemize}

\noindent
In the interest of ensuring the reproducibility of our results,
we also provide access to the source code used to generate the
data in this paper. See the section on open science on the next
page for details.
\vfill~

\clearpage

\section*{Declarations} %

\subsection*{Acknowledgements}

We thank Peiwen Ren and Lorenzo Contento for their advice on computing the lower
convex envelope of a surface; Nathan van Zee for his cartoon of a vitrimer
melt; and our colleagues from the ReBond network for fruitful discussions.

\subsection*{Funding}

This work was funded by the European Union under a Horizon~2020 grant [GA\,101119786].
Simulations were carried out on the \enquote{IBiSCo} high-performance
computing cluster, funded by a PON~2014--2020 grant [PIR\,01\_00011].

\subsection*{Open science and open access}

Following ReBond's commitment to open science, the code used to generate the
data for the figures in this paper is made publicly accessible in the
\emph{\foreignlanguage{french}{Université catholique de Louvain}}'s dataverse
for the ReBond project \autocite{rebond_dataverse}.

\begin{center}
	\url{https://dataverse.uclouvain.be/dataverse/rebond}
\end{center}

The Monte-Carlo bond-exchange logic has also been made available as an
open-source \href{https://pypi.org/project/MC-exchange/}{PyPI package}
installable through \verb|pip| under the name \verb|MC-exchange|. The source
code can be found on GitHub \autocite{magyari2026}.

\begin{center}
	\url{https://github.com/balintmagyari/MC_exchange}
\end{center}

For the purpose of open access, the authors have applied a Creative Commons
Attribution (CC~BY) licence to any Accepted Manuscript arising from this
submission.

\subsection*{Conflicts of interest}

The authors declare no competing financial interest.

\subsection*{LLM transparency statement}

A large language model (Google Gemini) was used by {B.\,M.}\ to assist in refining the Python scripts used for data analysis and to edit the manuscript for language and clarity. The authors rigorously reviewed and verified all {LLM}-generated outputs to ensure scientific accuracy and assume full responsibility for the final content.

\subsection*{Author contributions}

\textbf{A.\,A.\,R.\,C.} ---
	theoretical framework, manuscript writing
	(\cref{sec:intro,sec:theory,sec:discussion,sec:conclusion}) \\
\textbf{B.\,M.} ---
	simulations, data analysis, manuscript writing
	(\cref{sec:simulation,sec:conclusion}) \\
\textbf{G.\,I.} --- \textbf{E.\,v.\,R.} --- \textbf{D.~J.~R.} ---
	supervision, guidance, manuscript review and editing

\clearpage

~\vfill\tableofcontents\vfill
\newpage
~\vfill\listoffigures\vfill\listoftables\vfill~

\clearpage
\printbibliography

\clearpage
\appendix
\pagenumbering{roman} \setcounter{page}{1}
\renewcommand{\thefigure}{\Roman{figure}} \setcounter{figure}{0}
\renewcommand{\thetable}{\Roman{table}} \setcounter{table}{0}
\renewcommand{\theequation}{\roman{equation}} \setcounter{equation}{0}
\title{Supplementary Information for:\\ Entropic Phase Separation in Vitrimer--Polymer Melts}
\maketitle

\section{Full free-energy derivation} \label{apx:fullderivation}

In this section, we provide a derivation for the free energy of the system when the assumption of \enquote{phantom} cross-linker and stopper moieties is relaxed. For this fuller derivation, we will use a lattice argument on the four species comprising our system: \begin{enumerate}[(A)]
    \item the homopolymer matrix, consisting of \(n_\A\) chains of size \(N_\A\);
    \item the vitrimer backbone, similarly consisting of \(n_\B\) chains of size \(N_\B\);
    \item the (difunctional) cross-linker moiety, \(n_\C\) units of size \(N_\C\), connecting distinct stickers; 
    \item and the (monofunctional) stopper moiety (\(n_\D, N_\D\)), occupying any sticker not cross-linked. 
\end{enumerate} We will begin with the usual Flory--Huggins entropy of mixing for these four species, then extend it to the cross-linked case. Note that we assume that cross-linker and stopper molecules are always attached to a sticker of the vitrimers; no sticker site is left unfilled. This assumption of stoichiometry reflects the structure of stably synthesised vitrimers \autocite{rottger2017}. 

Imagine filling up a lattice with molecules of a given species \(i\); say the lattice has \(n_0\) sites and a coordination number \(z\). For the first molecule, occupying \(N_i\) sites, there are \(n_0\) choices for the first occupied lattice site, \(z\) for the next, some number \(y\) --- depending on the molecule's flexibility --- for the next, \(y\) again for the next, and so on until the end of the molecule. This molecule thus contributes \(n_0 z y^{N_i-2} \eqcolon n_0 \zeta_i\) choices to the total configuration count. (We wrap several constants into the species-dependent flexibility factor \(\zeta_i\); this allows us to do away with special treatment of the cases \(N_i=1\) and \(N_i=2\).) We can repeat this procedure for all the other molecules, with one modification: as we progressively fill up the lattice, some of the lattice sites will be already occupied. We thus weight every new choice by the probability that the site is free at that step. Repeating for all molecules \(n_i\) of a given species~\(i\) --- yielding a factor \((n_0\zeta_i)^{n_i}\) --- and for all species in the mixture, we obtain a configuration count of \begin{equation}
    \Omega = \prod_{\mathclap{\text{species }i}} {(n_0 \zeta_i)^{n_i}} \times \prod_{t=0}^{{n_0-1}} \mathbb{P}(\text{next site free after }t\text{ steps}) \,.
\end{equation} Following the classic mean-field derivation of \textcite[§\,\RN{12}-1a]{flory1953}, we set the probability that the next site is free to be the number of free sites after \(t\)~steps, \ie, \(\mathbb{P}(t) = (n_0 - t)/n_0\). Simplifying and dividing by \(n_i!\) for the Gibbs correction (indistinguishability within a species), we find \begin{equation}
    \Omega = \frac{n_0!}{n_0^{n_0}} \times \prod_{\mathclap{\text{species }i}} \frac{(n_0 \zeta_i)^{n_i}}{n_i!} \,.
\end{equation} This is the configuration count for the Flory--Huggins mixture theory. 

From this, we obtain the count for the cross-linked system by \begin{enumerate*}[(i)]
    \item allowing for all possible pairings of stickers,\label{blt:stickerpermutations}%
    \item removing degrees of freedom due to moieties being attached, and\label{blt:moietiesattached}%
    \item claiming that the orientation of the cross-linker is immaterial.\label{blt:crosslinkersymmetric}%
\end{enumerate*} \Cref{blt:crosslinkersymmetric} is easy, as it simply introduces \(n_\C\) factors of \(1/2\) into the overall count. \Cref{blt:moietiesattached,blt:stickerpermutations} are a bit more subtle, and require some care. Here we follow the argument of \textcite{dobrynin2004} to compute the number of pairings and the loss of degrees of freedom. Since we have already corrected for the indistinguishability, we treat the C- and D-moieties as ordered in what follows.

We first count the cross-linked sticker pairs, then include the stoppers. Choose a first sticker (\(sn_\B\) choices). The probability that the first cross-linker moiety starts on any of the \(x\) bondable sites adjacent to the first sticker is \(x/n_0\), as the moiety that could previously explore the whole volume is now restricted to lie on one of \(x\) sites. Here, \(x\) acts like a discretised bond volume. Then, the probability that any of the \(x\) sites on the other end of the C-moiety is occupied by another sticker is, by the mean-field approximation, \(x\times(sn_\B-1)/n_0\). And this is one pair. Then we repeat: pick a sticker (\(sn_\B-2\) choices), ask after the probability that it connects to the next C-moiety \((x/n_0)\), take the probability of seeing a new sticker \(x \times (sn_\B-3)/n_0\), repeat. After placing all C-moieties, we then pick our next new sticker out of a pool of \(sn_\B-2n_\C\) stickers, and note that the probability of seeing the first D-moiety is \(x/n_0\). Continuing this procedure until we have placed all D-moieties (thus covering all stickers), we obtain the product \begin{multline*}
    sn_\B \times x \times \frac{(sn_\B-1)}{n_0} \times \cdots \times (sn_\B-2n_\C+1) \times x \times \frac{(sn_\B-2n_\C)}{n_0} \\ 
    \times (sn_\B-[2n_\C-1]) \times \frac{x}{n_0} \times \cdots \times 1 \times \frac{x}{n_0} \,,
\end{multline*} which collapses to \((sn_\B)! \times \left({x}/{n_0}\right)^{sn_\B}\) (where we have used the assumption of stoichiometry \(sn_\B=2n_\C+n_\D\) to simplify the exponent). Thus, putting everything together, we obtain an expression for the configuration count \begin{equation}
    \Omega = \frac{n_0!}{n_0^{n_0}} \frac{(sn_\B)!}{2^{n_\C}} \left(\frac{x}{n_0}\right)^{s n_\B} \prod_{i=\A}^{\D} \frac{(n_0\zeta_i)^{n_i}}{n_i!} \,.
\end{equation} 
Taking the negative logarithm, applying Stirling's formula, and dividing by the number of lattice sites~\(n_0\), we obtain the dimensionless free energy per lattice site \begin{align}
 	f &\coloneq {(-TS)}/{(n_0k_\B T)} \notag \\
    &= 1 + \frac{\log2}{N_\C} \phi_\C - \frac{s\phi_\B}{N_\B} \log\left(\frac{xs\phi_\B}{\ec N_\B}\right) + \sum_i \frac{\phi_i}{N_i} \log\left(\frac{\phi_i}{\ec N_i \zeta_i}\right) \,,
\end{align} where the volume fraction of a given species is calculated 
according to \(\phi_i = N_i n_i/n_0\). Subtracting from the above 
expression any constants or terms linear in the concentration variables~\(\phi_i\), we 
obtain the standard form of our free energy, 
\begin{equation}\label{eq:apxsrf}
    f = - \frac{s\phi_\B}{N_\B}  \log\phi_\B + \sum_i \frac{\phi_i}{N_i} \log\phi_i \,.
\end{equation}
In these expressions, the volume fractions are related by the following system of equations: \begin{equation} \label{eq:volfracsyst}
	\begin{bmatrix}
		1 & 1 & 1 & 1 \\
		0 & sN_\B^{-1} & - 2N_\C^{-1} & - N_\D^{-1}
	\end{bmatrix}
	\begin{bmatrix}
		\phi_\A \\  \phi_\B \\ \phi_\C \\ \phi_\D
	\end{bmatrix}
	=
	\begin{bmatrix}
		1 \\ 0
	\end{bmatrix}
	\,.
\end{equation} A convenient choice of basis is the pair \((\phi,\psi)\) defined by \begin{equation}\label{eq:unphantomphipsi}
\phi =  1 - \phi_\A  \quad \text{and} \quad
\psi = \frac{2N_\B}{sN_\C} \, \phi_\C \,,
\end{equation} since \(\phi\) is the vitrimer volume fraction and both \(\phi\) and \(\psi\) are of comparable magnitude in the phase-space domain. (Indeed, the bounds \(0\leq \phi \leq 1\) and  \(0\leq \psi \leq (1 + s N_\C / 2 N_\B)^{-1} \,\phi\) form the boundary of the domain.) Note that, following the inclusion of the cross-linker and stopper moieties into the volume balance in \cref{eq:volfracsyst}, we lose the nice relationship \(p = \psi/\phi\), replacing it instead with \begin{equation}
	p = \frac{(N_\B+sN_\D) \, \psi}{N_\B \, \phi +   
		s(N_\D - N_\C/2) \, \psi} \,.
\end{equation}
Despite the increased complexity of this expression, lines of constant~\(p\) continue to be lines of constant ratio \(\psi/\phi\).

\subsection{Network cycles}\label{apx:cycles}

What part of the above argument suggests correct treatment of cycles? When we invoked the mean-field approximation, we claimed that the probability that a given site was occupied by a sticker was \((sn_\B - u)/n_0\), where \(u\) is the number of previously used-up stickers. Let us make a distinction: when we find a new sticker, it can either be on a cluster disconnected from the first sticker (with probability \(P_\text{dis}\)) or already connected to it through some path (with probability \(P_\text{con} = 1 - P_\text{dis}\)). In the case that the new sticker is completely disconnected, then it can explore the whole space, so the probability of seeing a sticker of this type is \(P_\text{dis} \times (sn_\B-u)/n_0\) as before. If instead it is somehow connected, it is constrained to some local neighbourhood~\(n_*\). Then the expected number of stickers in this neighbourhood is, \emph{by the mean-field approximation}, just the overall number of stickers reduced by the local neighbourhood size: \((sn_\B-u) \times n_*/n_0\). Hence the the probability of seeing a sticker of this second type is \(P_\text{con} \times (sn_\B-u)/n_* \times n_*/n_0\). Thus the total probability of seeing a sticker in the given site is \begin{equation*}
    \left( P_\text{dis} \times \frac{sn_\B-u}{n_0} \right) + \left( P_\text{con} \times \frac{sn_\B-u}{n_*} \times \frac{n_*}{n_0} \right) = \frac{sn_\B-u}{n_0} \,,
\end{equation*} exactly the expression we used before.  This is an established property of such models, stated \eg\ by \textcite[p.\,18]{tanaka2022}.

\subsection{Links to the same chain}

Technically, there is a slight contradiction, since the theory as stated should disallow stickers from connecting to stickers on the same chain. In practice, the error is vanishing. We compute it as follows: as we go along creating our network, each choice of sticker should exclude the \(s-1\) other stickers on the same chain. Thus we multiply the number of yet-unused stickers by the probability, for each sticker, that the chosen partner sticker is on a different chain: \[ \left(1-\frac{s-1}{sn_\B}\right)^{n_\C} \xrightarrow{n_\B \to \infty} \exp\left(-\frac{s-1}{2} \, p\right) \,.\] This is vanishingly small when we compute the intensive free energy \(f=-\log(\Omega)/n_0\).

\subsection{Nearest-neighbour correction}\label{apx:nn}

We provide also the nearest-neighbour correction for the four-species system. 
Since the derivation is essentially identical to that of \cref{eq:nnf} in the 
main body of the paper, we proceed in telegraphic style.
The entropic partition function is given by the expression
\begin{equation}
\begin{split}
	\Omega &= \frac{n_0!}{n_0^{n_0}} \, \left[\prod_{i} \frac{(n_0 \zeta_i)^{n_i}}{n_i!}\right] \times 
	 \mathbb{P}[{l}_{N_\B/s}]^{{\frac12}qs n_\B} \left[\frac{\left(s-\frac12 {qs} 
	 \right)!}{\left(\frac12 {qs} \right) ! \, (s-qs)!}\right]^{n_\B}  \\
	&\hspace{4em}\times \frac{(sn_\B-qsn_\B)!}{2^{(n_\C- \frac12 qs n_\B)}} \, \frac{n_\C!}{(n_\C- \frac12 qs n_\B)!}\, \left(\frac{x}{n_0}\right)^{(sn_\B- \frac12 qs n_\B)} \,,
\end{split}
\end{equation} recalling that \(n_0 = \sum_i N_i n_i\) and \(sn_\B = 2n_\C + n_\D\). In these expressions, the index \(i\) runs over all four moieties A, B, C, and D\@. Taking the logarithm and minimising over \(q\) gives the intensive free energy \begin{equation}
\begin{split}
f &=  
- \mathfrak{H}\!\left(\frac{s\phi_\B}{N_\B} - \frac qp \frac{\phi_\C}{N_\C}\right) 
+ \sum_i \frac{\mathfrak{H}(\phi_i)}{N_i}\\
&\hspace{4em}+ \left[\mathfrak{H}(1-q/p) + \mathfrak{H}(q/p) + \frac qp \log\left(\frac{2N_\B}{\ec s k}\right)\right] \frac{\phi_\C}{N_\C}  \,,
\end{split}
\end{equation}
with \(\mathfrak{H}(z) \coloneq z \log(z)\) and \(q \approx kp / (k+2\phi_\B)\). For numerical work, it is convenient to re-write this in terms of \((\phi,\psi)\) as defined in \cref{eq:unphantomphipsi}, and use the rescaled free energy \(N_\B f/s\).

\section{Energetic interactions}\label{apx:chi}

It is rare in real melts to find a system which is truly in theta-solvent conditions. What consequence does adding an interaction energy between monomers of different species bring? 
While the true effect of the backbone chemistry on vitrimer mechanics is still unclear \autocite{ge2025}, we believe the exploration of a toy model, based on the Flory--Huggins interaction parameter, is helpful as a first approach to the subtleties of including chemistry into our predictions. 

\begin{figure}[bt]
	\centering
	\pgfplotsset{width=0.6\textwidth}
	\includegraphics{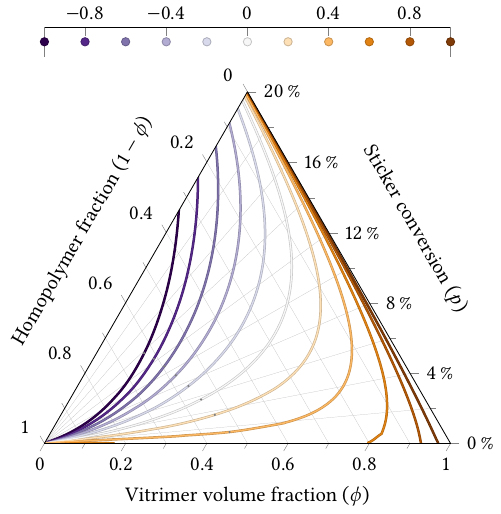}
	\caption[Effect of an energetic interaction]{Effect of an energetic 
	interaction (\(\degreepolym \A = \degreepolym \B = 500\), \(s=50\), \(k 
    \approx 0.30\); \(\chi\) varying). \emph{The Flory 
	interaction parameter between backbones and dioxaborolanes controls 
	the position of the binodal line.}}
	\label{fig:chiphasediagram}
\end{figure}

\subsection{A first look}

It is in the non-phantom framework --- where all four moieties have 
independent existence --- that the inclusion of an energetic interaction 
is most meaningful. In the abstract, one could provide interactions 
between all \(4(4-1)/2=6\) possible combinations of moiety. In a concrete 
example, however, like that of a dioxaborolane-based 
stopper--cross-linker pair, that number halves. If, furthermore, the 
vitrimer is of the same backbone as the polymer matrix, the simplest 
correction to the free energy might to be of the form \(\Chg f = \chi 
(\phi_\A + \phi_\B)(\phi_\C + \phi_\D) \,,\) namely as a Flory 
interaction parameter~\(\chi\) taken between unlike chemical species. 
This is the most sensible expression in the limit as the inter-function 
spacing \(\degreepolym \B/s\) grows, since then contacts between 
non-identical chemical elements become increasingly rare. Using the 
expression \(\chi \phi_\A (1-\phi_\A)\) --- which sets homopolymer chains 
against all components of the vitrimer --- while seeming natural at 
first, would vastly overestimate the enthalpic repulsion when 
\(\degreepolym \B/s\) is large. The interaction parameter~\(\chi\) will have 
an effect on the binodal when it is comparable in size to the stable, 
Flory--Huggins part of the free energy expression. Since \(\phi_\C \sim 
s\phi_\B/N_\B\), and the stable term in \cref{eq:apxsrf} scales as 
\(1/N_\B\) (assuming \(N_\C\) and \(N_\D\) are roughly unity, and \(N_\A 
\sim N_\B\)), we expect this to occur when \(s\chi \sim 1\).

Implementing \(\Chg f\) for \(\chi\in\{1,0.8,\dots,-1\}\)
gives the diagrams in \cref{fig:chiphasediagram}. (Parameter values 
\(\degreepolym i = 500\), \(s=50\) were chosen to keep algorithm run-time
low.) We see that the effect of the interaction parameter is fairly 
straightforward: the binodal advances for \(\chi>0\) and recedes for 
\(\chi<0\). The natural scale which varies the binodal, \(\chi\sim 
10^{-1}\), is consistent with the interpretation that the relevant 
scaling for the interaction parameter is \(\chi \sim 1/s\).

\subsection{Splitting the molecules}%

\begin{figure}[tb]
	\centering
	\includegraphics[width=0.6\linewidth]{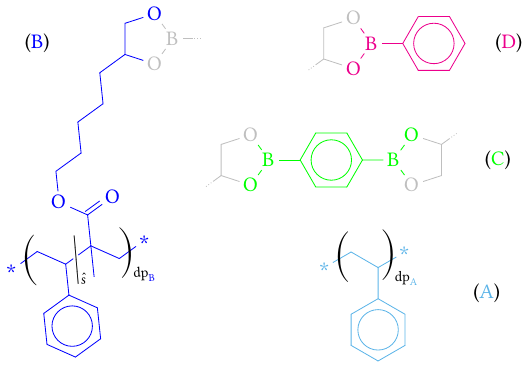}
	\caption[Polystyrene/dioxaborolane vitrimer melt]{Chemistry of a polystyrene/dioxaborolane vitrimer melt.}
	\label{fig:PS2}
\end{figure}

\begin{figure}[thb]
	\centering
	\includegraphics[width=0.6\linewidth]{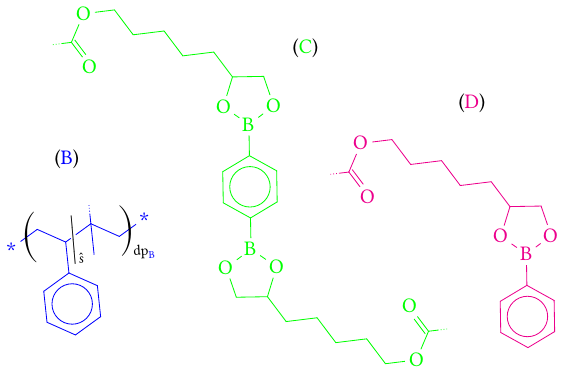}
	\caption[Variant segmentation of polystyrene/dioxaborolane vitrimer]{Alternative method of chopping up a polystyrene vitrimer with dioxaborolane functions into species B, C and~D.}
	\label{fig:PSvar}
\end{figure}

\Cref{fig:PS2} shows the chemical structure of a polystyrene-based vitrimer--homopolymer melt with dioxaborolane-based cross-linker and stopper, categorised into the four species A, B, C and D. The cutting-point between B and C/D is chosen to be the point where the chemical bonds break during a cross-link exchange interaction. 
We note, however, that splitting C/D from B at a different point than that indicated in \cref{fig:PS2} may allow the expression \(\chi (\phi_\A+\phi_\B)(\phi_\C+\phi_\D)\) better to capture the enthalpic behaviour of these materials. An example alternative split is shown in \cref{fig:PSvar}. The choice of optimal split can only be made with a detailed understanding of the chemistry involved.

Incidentally, a split can always be chosen such that \(N_\B = \degreepolym \B\) 
(leaving, in our example material, \(N_\C = 3 +{3}/{52}\) 
and \(N_\D = 2 - {5}/{52}\)),
which does simplify some algebra, but is unlikely to correspond 
to anything physically meaningful in terms of the chemistry.

\subsection{Globules and micro-phase segregation}

The true effect of an energetic interaction, however, is likely to be 
more complicated than the simple addition of a term to the free energy. 
In the presence of a repulsive interaction between dioxaborolane and 
polystyrene, for example, we would expect the formation of globule-like 
structures or micro-phase separation in the vitrimer chains --- 
especially when the vitrimer is dilute in the polymer matrix, when also 
assumptions of the mean-field theory begin to weaken. The free energy in 
these regimes then needs to be treated differently, for example by 
assuming from the outset the formation of globular structures for 
vitrimer chains in the derivation of the free energy 
\autocite{semenov1998,dobrynin2004}. As concerns micro-phase separation, 
however, this is likely beyond the predictive power of a theory such as this, 
which is based entirely on the assumption of a mean field.

\section{Cross-linker of arbitrary functionality}\label{apx:funrxlk}

In this section, we provide a calculation of the system entropy when the cross-linker is not difunctional, but has an arbitrary functionality \(r\geq2\). For simplicity, we will return to the \enquote{phantom} approximation, where cross-linker and stopper do not occupy any physical volume. The calculation is a trivial generalisation of the difunctional case when cross-links are only between different chains, but requires some care when cross-links are allowed to involve stickers from the same chain. We will again approximate the same-chain entropy with its leading-order contribution, namely when all local bonds are between runs of successive (\enquote{nearest-neighbour}) stickers. Longer-range bonds are suppressed polynomially, scaling as the inter-sticker distance to the power of \(-3/2\).

\WaOignorenext%
Define \(q_i\) as the proportion of stickers involved in a run of~\(i\) successive stickers on one chain all attached to the same cross-linker. Then defining \begin{equation*}
	q \coloneq \sum_{i=2}^{r} q_i \quad \text{and} \quad \bar q \coloneq \sum_{i=2}^r \frac1i q_i \,,
\end{equation*} we use a \enquote{stars-and-bars}--type argument within each 
vitrimer chain with \(r-1 = |\{2,\ldots,r\}|\) different types of \enquote{bars}, 
each occurring respectively \(q_i/i\) times in the chain. The total number of 
objects (\enquote{stars} plus \enquote{bars}) is \((s-qs) + \bar qs\), giving a 
combinatorial factor \begin{equation}\label{eq:funrxlk1}
 \frac{[s(1-q+\bar q)]!}{[s(1-q)]! \, \prod_{i=2}^{r} \left(\frac{1}{i}sq_i\right)!} \,.
\end{equation} This configuration occurs with probability 
\(\mathbb{P}[{l}_{N/s}]^{s(q-\bar q)}\), where the exponent is the number of 
loops formed by attachment to the cross-linker (namely, \(i-1\) for every 
cross-linker combining \(i\)~successive stickers).

We then have to place all remaining stickers. We use a similar picture to the one in \cref{sec:sr}, but with more types of cross-linker: in brackets those fully free, and in parentheses those already partially attached due to nearest-neighbour cross-linking. The total number of asterisks to place is \(sn(1-q)\).
\begin{equation*}
	 \overbrace{()()\cdots()}^{n\frac{sq_r}{r}} \ds
	 \overbrace{(*)(*)\cdots(*)}^{n\frac{sq_{r-1}}{r-1}} \ds
	 \cdots \ds
	 \overbrace{(\underbrace{*,\dots,*}_{r-2})\cdots(*,\dots,*)}^{n\frac{sq_{2}}{2} \text{ have }r-2\text{ sites free}}
	 \ds\bigg|\ds 
	 \overbrace{[\underbrace{*,\dots,*}_{r}]\cdots[*,\dots,*]}^{sn(\frac pr - \bar q)\text{ non-local}} 
	 \ds\bigg|\ds 
	 \overbrace{* \,* \, \cdots \, *\vphantom{()}}^{sn(1-p)\text{ free}} 
\end{equation*} This picture allows us to immediately write down the remaining combinatorial factor, \begin{equation}\label{eq:funrxlk2}
 \frac{1}{r!^{sn(\frac 1 r p - \bar q)} \, \prod_{i=2}^{r} (r-i)!^{s n q_i/i}} \times \frac{[sn(1-q)]!}{[sn(1-p)]! \, [sn(\frac1r p - \bar q)]!} 
\end{equation} where we have divided by permutations of stickers within parentheses or brackets, along with permutations of the brackets or of the free stickers. We have \emph{not} divided by permutations of the parentheses, as these are not interchangeable: a cross-linker combining \(i\)~nearest-neighbour stickers on one chain is distinct from a cross-linker combining~\(i\) on another. The probabilistic weight associated with \cref{eq:funrxlk2} has exponent given by the number of bound asterisks in the picture, yielding \((x/n_0)^{sn((1-\frac1r)p + \bar q - q)}\). Combining together \cref{eq:funrxlk1,eq:funrxlk2}, we find a partition function \begin{equation}\label{eq:funrxlkOmega} 
	\begin{split}
		\frac{\Omega}{\Omega_\text{FH}} &=  \frac{\left.[s(1-q+\bar q)]!\right.^n \, \left(\frac14 k x s / N\right)^{sn(q-\bar q)}}{\left.[s(1-q)]!\right.^n \, \prod\limits_{i=2}^{r} \left.({sq_i}/{i})!\right.^n} \\
	 &\qquad \times 
	 	\frac{[sn(1-q)]! \, \left(\frac{x}{n_0}\right)^{sn[(1-\frac1r)p-q+\bar 
	 	q]}}{r{!}^{sn(\frac 1r p - \bar q)} \, [sn(1-p)]! \, [sn(\frac1rp - \bar q)]!
        \prod\limits_{i=2}^{r} 
	 	{(r-i)!}^{{s n q_i}/i}
        } 
	\end{split}
\end{equation} where we have substituted \(\mathbb{P}[{l}_{N/s}] = \frac14 k x s 
/N\).

Nearest-neighbour cross-links with runs of more than two elements are 
exponentially suppressed as they involve the factor \(\mathbb{P}[{l}_{N/s}]\) 
raised to the number of loops. Therefore, we approximate \cref{eq:funrxlkOmega} 
with its leading-order behaviour, namely where \(q_i = 0 \  \forall i>2\). 
Performing that simplification and computing the free energy gives the expression 
\begin{equation}\label{eq:funrxlkf}
\begin{split}
	\frac Ns f &= \frac1s \left(\mathfrak{H}(\phi)+\frac NM \mathfrak{H}(1-\phi)\right) + \mathfrak{H}(\phi-\psi) \\
	&\qquad + \frac12 \mathfrak{H}\!\left(\frac2r \psi-q\phi\right) - \mathfrak{H}\!\left([1-\frac 12q] \,\phi\right) \\
	&\qquad\qquad+ \frac12 \mathfrak{H}(q\phi) +  \frac{1}{2} q\phi \log\left(\frac{4/\ec}{kr(r-1)}\right) 
\end{split}
\end{equation} as required. This is minimised when \begin{equation}\label{eq:rfuncq}
	\frac{(2 - q)q}{2p-rq} = (r-1)\frac{k}{2\phi}
\end{equation} leading to the linearisation \begin{equation*}
	q \approx \frac{2(r-1)kp}{r(r-1)k + 4\phi} \,.
\end{equation*} Inspection shows that small, reasonable values of the cross-linker functionality \((r \lesssim 10)\) give near-identical expressions for the free energy in \cref{eq:funrxlkf}  equivalent to rescaling \(\phi\mapsto\phi/(r-1)\), and are thence not expected to yield qualitatively different results for the bulk of the phase diagram. 

In contrast with the case \(r=2\), however, \cref{eq:rfuncq} for general \(r\) does \emph{not} yield the correct behaviour \(q\to p\) as \(\phi\to0\) (dilute vitrimer limit). This is because cross-linking runs of more than two successive stickers gain in importance when associations are primarily intra-chain, and need to be added back into the model, reversing the approximation immediately preceding \cref{eq:funrxlkf}.

\section{Additional plots}\label{apx:plots}

\begin{figure}[p]
    \centering
	\includegraphics[height=0.89\textheight]{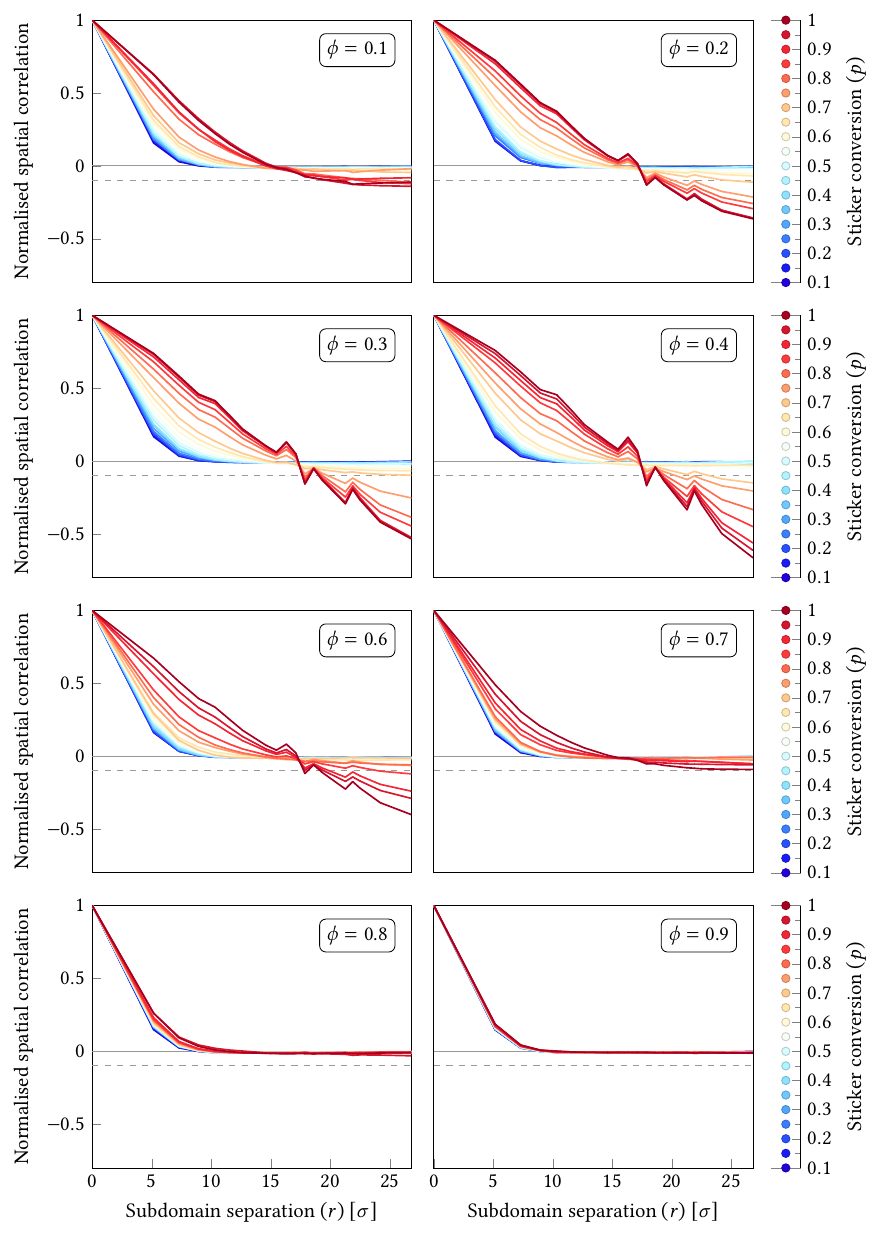}
    \caption[Further spatial correlation]{Normalised spatial correlation as a function of subdomain separation for \(\phi\) between \(0.1\) and \(0.9\) excluding \(0.5\) (for which see \cref{fig:spatial_corr}).}
    \label{fig:apx_spatial_corr}
\end{figure}

\begin{figure}[p]
    \begin{center}
        \includegraphics[width=0.89\textwidth]{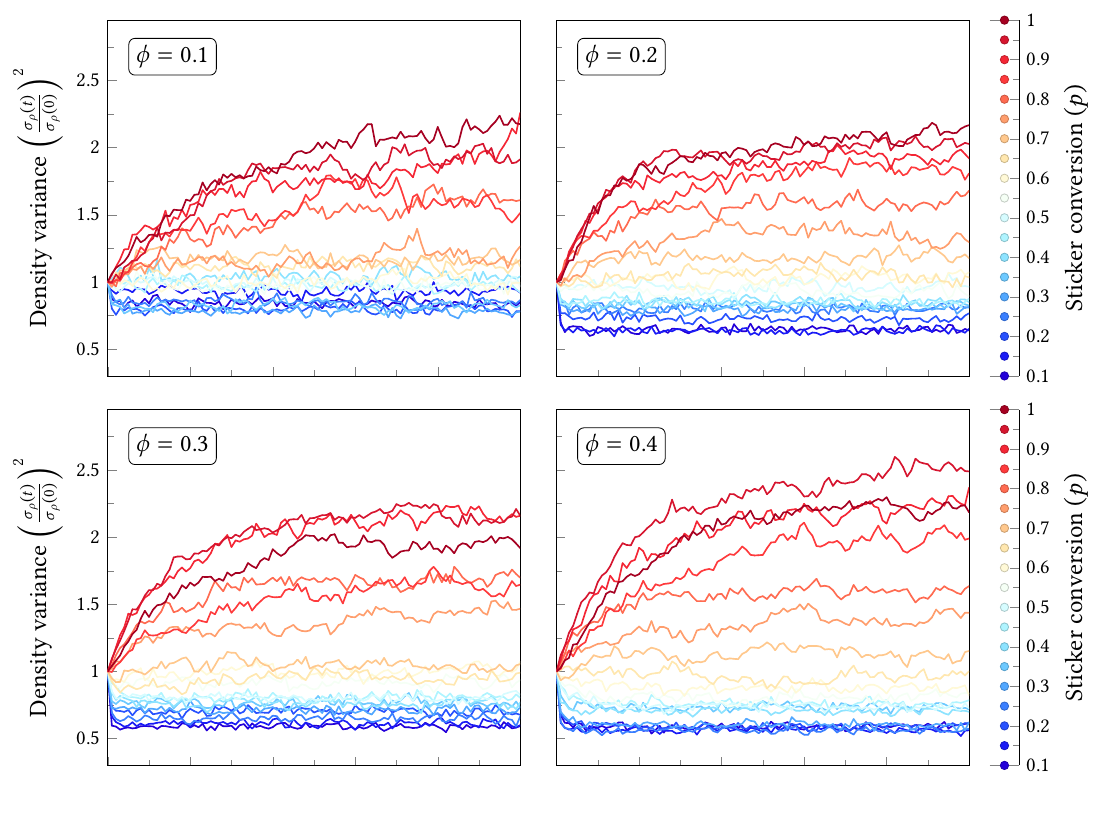}
        
        \vspace{-1.7em}
        
        \includegraphics[width=0.89\textwidth]{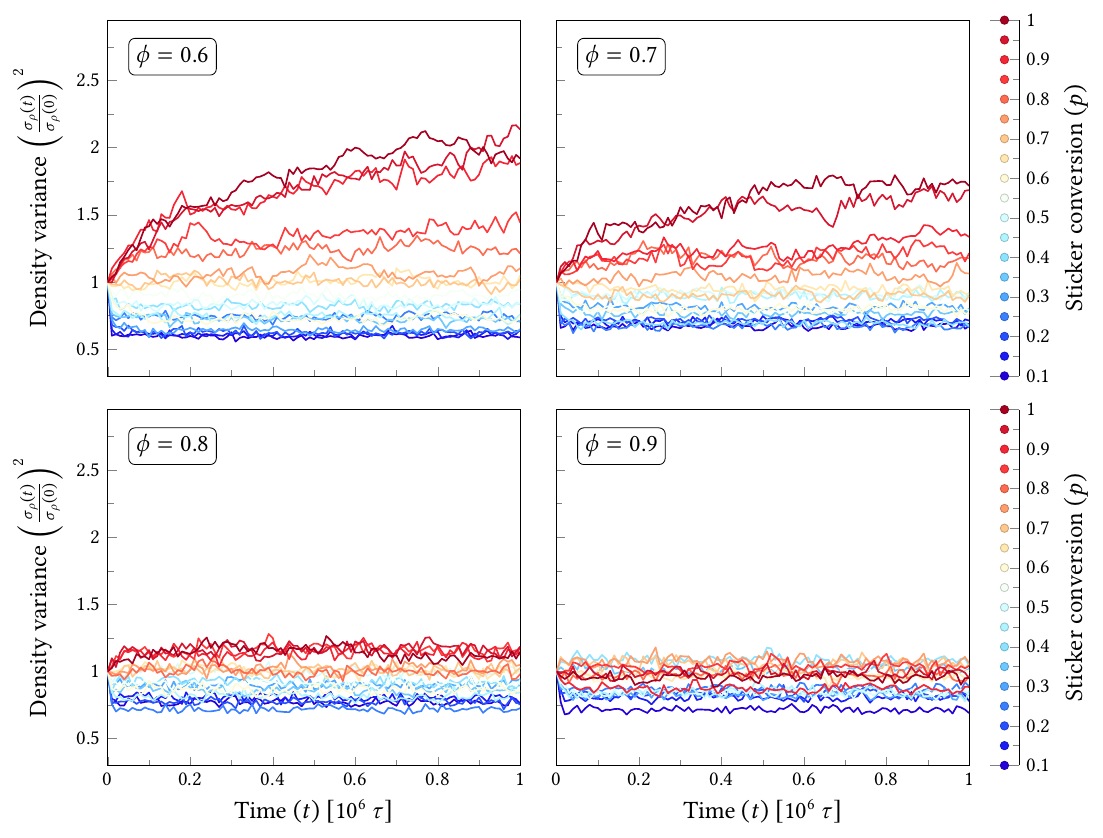}

        \vspace{-1.2em}
    \end{center}
    \caption[Further density variance]{Normalised density variance as a function of simulation time for \(\phi\) between \(0.1\) and \(0.9\) excluding \(0.5\) (for which see \cref{fig:density_var}).}
    \label{fig:apx_density_var}
\end{figure}

\begin{figure}[p]
    \centering
	\includegraphics[height=0.89\textheight]{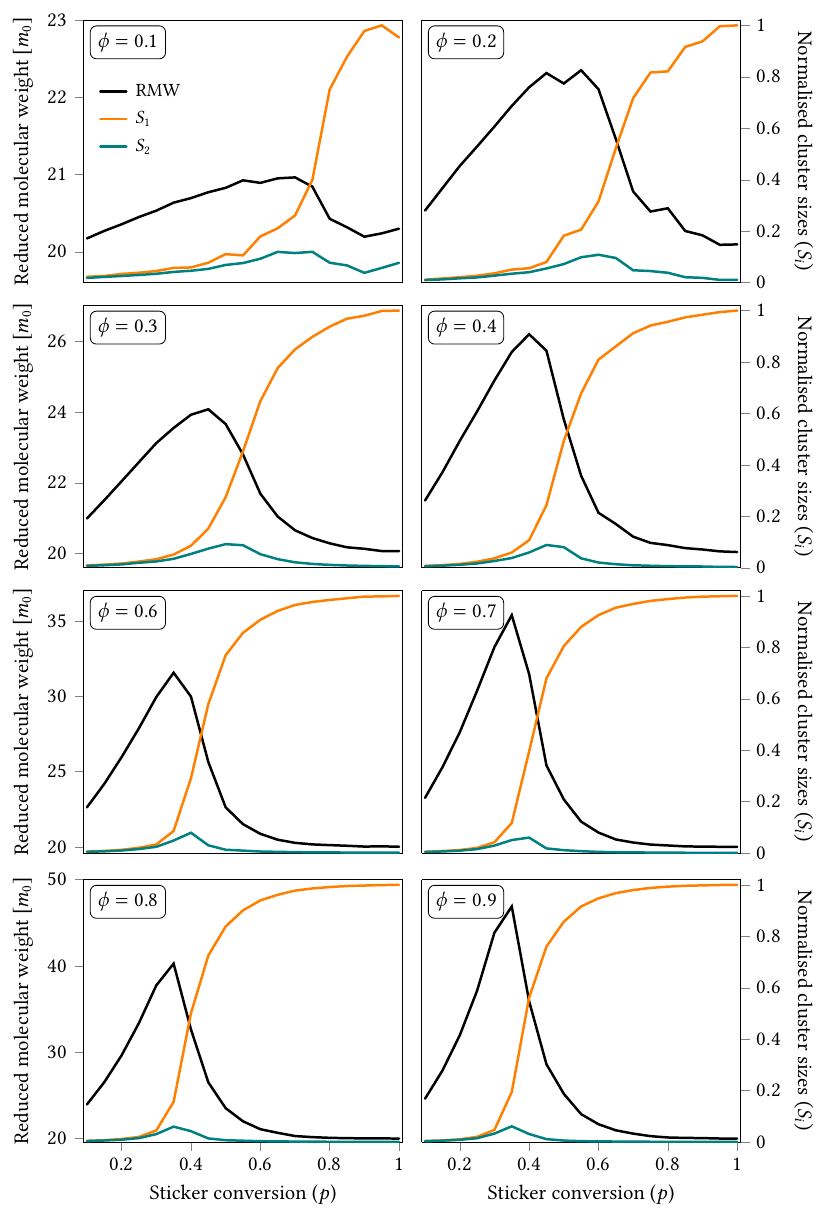}
    \caption[Further percolation transition measurements]{Data required for the measurement of the percolation transition, for \(\phi\) between \(0.1\) and \(0.9\) excluding \(0.5\) (for which see \cref{fig:percolation}).}
    \label{fig:apx_percolation}
\end{figure}

\Cref{fig:spatial_corr,fig:density_var,fig:percolation} in the main text are for \(\phi=0.5\). The corresponding plots for all remaining simulated values of~\(\phi\) are given in this section, in \cref{fig:apx_spatial_corr,fig:apx_density_var,fig:apx_percolation} respectively. These show: the normalised spatial correlation functions (\cref{fig:spatial_corr,fig:apx_spatial_corr}), the normalised density variance (\cref{fig:density_var,fig:apx_density_var}), and the data for the measurement of the percolation transition (\cref{fig:percolation,fig:apx_percolation}).

\clearpage
\section{Asymptotic calculation of the plait point}\label{apx:pp-asymptotics}

\DIFadd{
This section provides the justification for the position of the plait point as~\(s\to\infty\), represented by \cref{eq:pp-scaling,eq:pp-scaling-phi,eq:pp-scaling-psi} in the main text. As the full model with steric cross-linker and stopper is analytically intractable, the asymptotic analysis is based on the simplified model with \enquote{phantom} cross-links. For ease of reference, we reproduce here the free energy in the phantom model without the nearest-neighbour correction, \cref{eq:srf}, as well as the conditions satisfied at the plait point, \cref{eq:spinodal,eq:plait}.
\begin{equation}\tag{\ref{eq:srf}}
    \frac Ns f \approx \frac1s \left[\mathfrak{H}(\phi) + \frac{N}{M} \mathfrak{H}(1-\phi)\right] 
    + \mathfrak{H}(\phi-\psi) + \frac12 \mathfrak{H}(\psi) - \mathfrak{H}(\phi) \,, \quad\mathfrak{H}(x)\coloneq x\log x
\end{equation} 
Recalling the notation \(f_{,xyz} \coloneq \partial^3f/\partial x \partial y \partial z\), the equations we are required to solve are
\begin{equation}\tag{\ref{eq:spinodal}}
    f_{,\phi\phi}f_{,\psi\psi} -(f_{,\phi\psi})^2 = 0 \,, \quad \text{and}
\end{equation} 
\begin{equation}\tag{\ref{eq:plait}}
    f_{,\psi\psi\psi} f_{,\phi\psi} f_{,\phi\phi}
    - 3 (f_{,\phi\psi})^2 f_{,\phi\psi\psi}
    + 3 f_{,\phi\psi} f_{,\psi\psi} f_{,\phi\phi\psi}
    - (f_{,\psi\psi})^2 f_{,\phi\phi\phi}
    = 0  \,.
\end{equation} In the interest of reproducibility, we also provide at the end of this section a print-out of the Wolfram \emph{Mathematica} notebook used to perform this calculation.

    Finding the asymptotic behaviour of the plait point as \(s\to\infty\) requires simultaneously solving \cref{eq:spinodal,eq:plait} to first order in \(1/s\). For
    a simple case with no nearest-neighbour cross-linking and \(N=M\), putting \cref{eq:srf,eq:spinodal,eq:plait} into a computer algebra system provides an exact solution for the plait point, which can be asymptotically expanded to \begin{gather}
        \phi = \frac12 - \frac3{2s} + \bigO\left(s^{-2}\right)
        \ds\text{and}\ds
        \psi = \frac1s + \bigO\left(s^{-2}\right)
        \notag \\
        \implies p \coloneq \frac\psi\phi = \frac2s + \bigO\left(s^{-2}\right) \,.
        \label{eq:pp-scaling-k=0-N=M}
    \end{gather} Already this tells us that in this simple case, a rule of thumb for the transition into immiscibility is that two stickers per chain be cross-linked to other chains. 
    The remainder of this section seeks to generalise the result to \(N\neq M\) and to the more refined free energy of \cref{eq:nnf} ---
    \begin{equation}
        \begin{split}
            \frac Ns f &= \frac1s \left[\mathfrak{H}(\phi)+\frac NM \mathfrak{H}(1-\phi)\right] +
            \mathfrak{H}(\phi-\psi) \\
            &\hspace{4em} + \frac12 \mathfrak{H}(\psi-q\phi) - \mathfrak{H}\!\left([1-\frac 12q]\,\phi\right) \\
            &\hspace{8em} + \frac12 \mathfrak{H}(q\phi) +  \frac{1}{2} q\phi \log\left(\frac{2}{k\ec}\right) 
        \end{split}\tag{\ref{eq:nnf}}
    \end{equation} --- with the small-\(p\) approximation \(q\approx kp/(k+2\phi)\) for the sticker fraction of nearest-neighbour cross-links.
    
    We attempt to find the solution by expanding in orders of~\(1/s\), and matching coefficients at every order. Unfortunately, \cref{eq:spinodal,eq:plait} appear to be singular in~\(N/M\), so the zeroth-order solution cannot be simply computed. 
    Previously, we found that~\(\psi\) vanished to zeroth order; we assume this to hold true even when some cross-links are lost to bonds within the same chain.
    Taking \(\psi\to0\) in~\cref{eq:nnf} and solving only \(f_{,\phi\phi\phi}=0\) --- corresponding to \cref{eq:plait} --- thus provides an ansatz
    \begin{equation}\label{eq:phi0}
    	\phi = \phi_0 \coloneq \frac1{1+\sqrt{N/M}}
    \end{equation}
    and~\(\psi = \psi_1/s\) for the leading-order solution. Using a computer algebra system, we substitute this ansatz into \cref{eq:spinodal} and solve for~\(\psi_1\) to leading order in~\(1/s\), yielding
    \begin{equation}\label{eq:psi1}
        \psi_1 \coloneq 1 + \frac k 2 \left(1 + \sqrt{\frac N M}\right) \,.
    \end{equation} The two terms~\(\phi_0\) and~\(\psi_1\) are enough to compute the number of stickers cross-linked to other chains at the plait point to leading order in~\(1/s\). This is
    \begin{equation}\label{eq:pp-scaling-apx}
        (p-q) \, s = \frac{\psi_1}{\phi_0} \left(1 - \frac{k}{k+2\phi_0}\right) + \bigO\left(\frac1s\right) = 1 + \sqrt{\frac NM} + \bigO\left(\frac1s\right) \,,
    \end{equation} consistent with \cref{eq:pp-scaling-k=0-N=M}.
    It can verified that the results of this scaling are compatible with \cref{fig:realphasediagram} in the main text, granting credence to the approximation of the cross-linker (and stopper) as point-like, \enquote{phantom} entities.
    
    For completeness, we calculate one more order. Trying the new ansatz
    \begin{equation*}
    	\phi = \phi_0 + \frac{\phi_1}{s} \quad \text{and} \quad
    	\psi = \frac{\psi_1}{s} + \frac{\psi_2}{s^2} \,,
    \end{equation*} with~\(\phi_0\) and~\(\psi_1\) given by \cref{eq:phi0,eq:psi1} respectively, we can simultaneously solve the leading orders of \cref{eq:spinodal,eq:plait} to give \begin{gather}
    \phi_1 \coloneq - \frac{3}{2} \sqrt{\frac{N}{M}} \qquad \text{and} \\[.5em]
    \psi_2 \coloneq \frac18 \left(1+\sqrt{\frac NM}\right)
    \left(8 + 4k + k^2 - 2 \, (4-k)(3+k)\sqrt{\frac NM} - (6-k)k \frac NM\right) \,.
    \end{gather} Note that the equation for \(\phi_1\) actually reduces to \(\phi_1^2 = 9N/4M\), leaving the sign of~\(\phi_1\) indeterminate; for consistency with \cref{eq:pp-scaling-k=0-N=M}, however, we must choose the negative branch \(\phi_1 < 0\).
}

\includepdf[pages=-,
            pagecommand={},
            ]{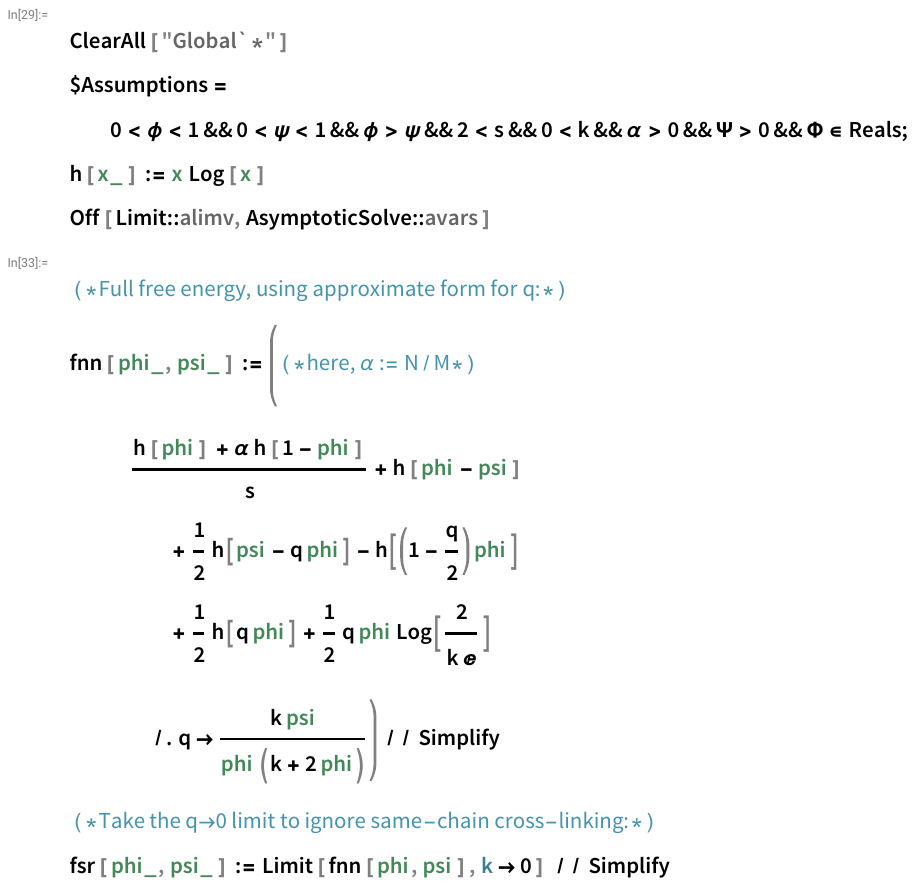}

\end{document}
\typeout{get arXiv to do 4 passes: Label(s) may have changed. Rerun}